\documentclass[10pt]{article}
\usepackage[scaled]{helvet}
\usepackage[T1]{fontenc}
\usepackage{apacite}
\usepackage{authblk}
\usepackage{caption}
\usepackage{color}
\usepackage{enumerate}
\usepackage{float}
\usepackage{graphicx}
\usepackage[colorlinks=true,citecolor=dblue,urlcolor=antiquefuchsia]{hyperref}		\hypersetup{linkcolor=black}
\usepackage[utf8]{inputenc}
\usepackage{natbib}
\usepackage{setspace}
\usepackage{fancyhdr}
\AtBeginDocument{%
    
}

\definecolor{antiquefuchsia}{rgb}{0.57, 0.36, 0.51}
\definecolor{dblue}{RGB}{6,69,173}

\begin{document}

\title{\huge \textbf{Computational modelling of angiogenesis: \\ The importance of cell rearrangements during vascular growth}}

\author[1,*]{Daria Stepanova}%
\author[2,3]{Helen M. Byrne}%
\author[2]{Philip K. Maini}%
\author[4,5,6]{Tom\'{a}s Alarc\'{o}n}%
\affil[1]{\normalsize{Laboratorio Subterr\'{a}neo de Canfranc, Canfranc-Estaci\'{o}n (Huesca), Spain}}%
\affil[2]{\normalsize{Wolfson Centre for Mathematical Biology, Mathematical Institute, University of Oxford, Oxford, UK}}%
\affil[3]{\normalsize{Ludwig Institute for Cancer Research, Nuffield Department of Medicine, University of Oxford, Oxford, UK}}%
\affil[4]{\normalsize{Instituci\'{o} Catalana de Recerca i Estudis Avan\c{c}ats, Barcelona, Spain}}%
\affil[5]{\normalsize{Centre de Recerca Matem\`{a}tica, Bellaterra (Barcelona), Spain}}%
\affil[6]{\normalsize{Departament de Matem\`{a}tiques, Universitat Aut\`{o}noma de Barcelona, Bellaterra (Barcelona), Spain}}%
\affil[*]{\normalsize{\href{mailto:dstepanova@lsc-canfranc.es}{\textcolor{antiquefuchsia}{dstepanova@lsc-canfranc.es}}}}%

\vspace{-1em}

\date{\today}
\begingroup
\let\center\flushleft
\let\endcenter\endflushleft
\maketitle
\endgroup

\begin{abstract}
\noindent Angiogenesis is the process wherein endothelial cells (ECs) form sprouts that elongate from the pre-existing vasculature to create new vascular networks. In addition to its essential role in normal development, angiogenesis plays a vital role in pathologies such as cancer, diabetes and atherosclerosis. Mathematical and computational modelling has contributed to unravelling its complexity. Many existing theoretical models of angiogenic sprouting are based on the `snail-trail' hypothesis. This framework assumes that leading ECs positioned at sprout tips migrate towards low-oxygen regions while other ECs in the sprout passively follow the leaders' trails and proliferate to maintain sprout integrity. However, experimental results indicate that, contrary to the snail-trail assumption, ECs exchange positions within developing vessels, and the elongation of sprouts is primarily driven by directed migration of ECs. The functional role of cell rearrangements remains unclear. This review of the theoretical modelling of angiogenesis is the first to focus on the phenomenon of cell mixing during early sprouting. We start by describing the biological processes that occur during early angiogenesis, such as phenotype specification, cell rearrangements and cell interactions with the microenvironment. Next, we provide an overview of various theoretical approaches that have been employed to model angiogenesis, with particular emphasis on recent \textit{in silico} models that account for the phenomenon of cell mixing. Finally, we discuss when cell mixing should be incorporated into theoretical models and what essential modelling components such models should include in order to investigate its functional role.

\vspace{5pt}
\noindent \textbf{This article is categorised under:}

\begin{quote}
Cardiovascular Diseases \textgreater{} Computational Models~

Cancer \textgreater{} Computational Models
\end{quote}

\vspace{5pt}
\noindent \textbf{Article Category}

\noindent Overview

\vspace{5pt}
\noindent \textbf{Keywords}

\noindent Angiogenesis, cell rearrangements, cell mixing, snail-trail model of angiogenesis, computational and mathematical modelling

\vspace{5pt}
\noindent \textbf{Funding information}

\noindent This work is supported by the Spanish State Research Agency (AEI) through the Severo Ochoa and Mar\'{i}a de Maeztu Program for Centers and Units of Excellence in R\&D (CEX2020-001084-M). We thank CERCA Programme/Generalitat de Catalunya for institutional support. DS and TA have been funded by grants RTI2018-098322-B-I00 and PID2021-127896OB-I00 funded by MCIN/AEI/10.13039/501100011033 ``ERDF A way of making
Europe''.

\par\null%
\end{abstract}

\newpage

\section{Introduction} \label{Section1}

\begin{figure}[b!]
\begin{center}
\includegraphics[width=0.955\columnwidth]{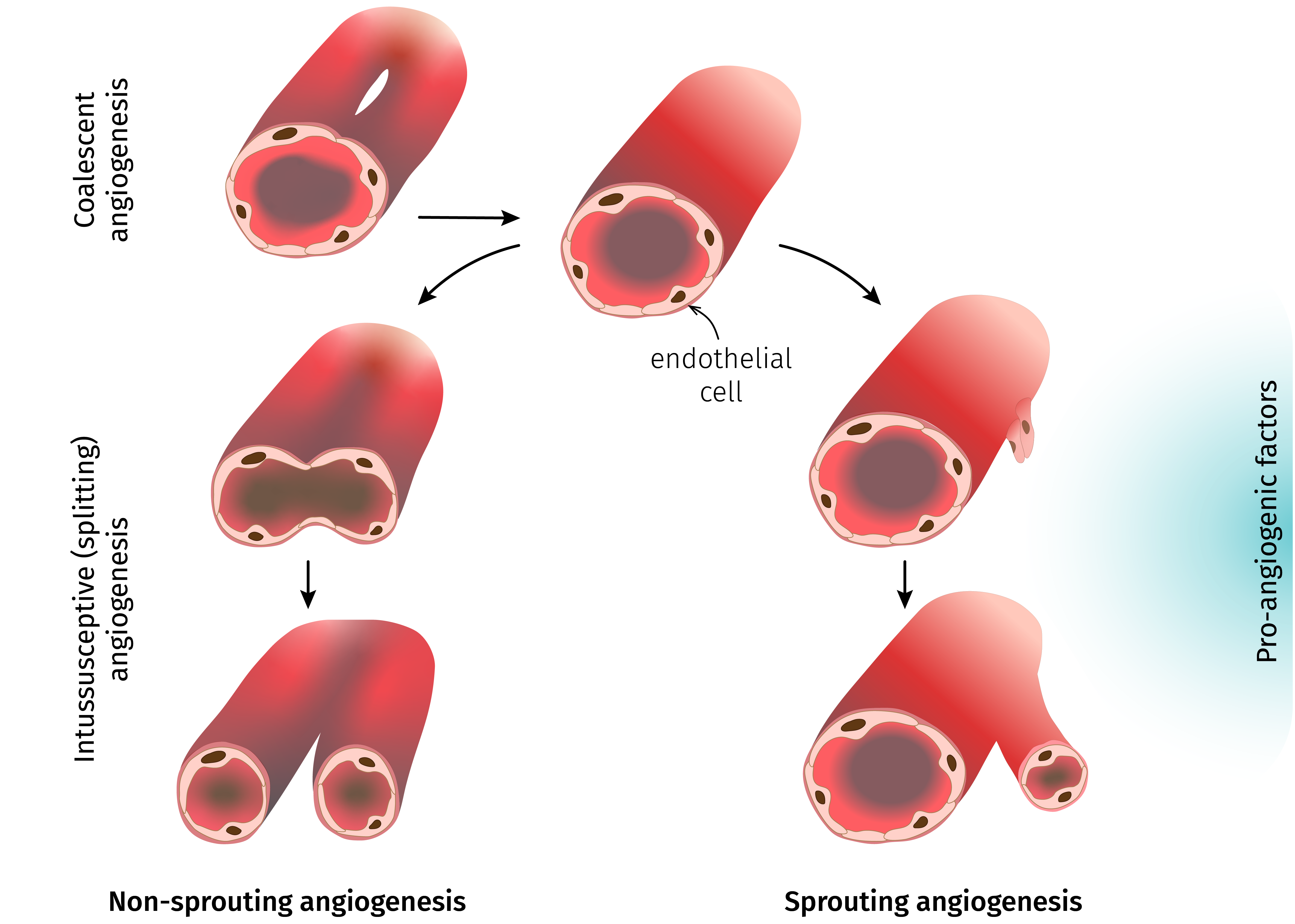}
\caption{\textbf{An illustration of non-sprouting and sprouting forms of angiogenesis.} The non-sprouting form of angiogenesis includes intussusceptive angiogenesis (depicted on the lower left) and coalescent angiogenesis (shown on the upper left). Vessel intussusception involves the splitting of a capillary into two neighbouring vessels \cite{Burri_2004,Mentzer_2014}. In contrast to vessel splitting, coalescence is the process by which capillaries fuse or coalesce to increase blood circulation \cite{Nitzsche2022coalescent}. Sprouting angiogenesis (depicted on the right) occurs when low-oxygen tissues release pro-angiogenic stimuli that activate ECs in the underlying vascular bed. This leads to the generation of new sprouts through the migration and proliferation of ECs towards the source of angiogenic cues.}
\label{Fig1}
\end{center}
\end{figure}

Blood vascular networks supply oxygen and nutrients and remove waste from the body’s tissues. Homeostasis in living systems relies on adequate formation and maintenance of the vasculature. In embryogenesis, endothelial progenitor cells assemble in cord-like structures to create a primitive vascular plexus. This process of \emph{de novo} formation of the initial vascular network is called vasculogenesis \cite{Eelen_2018}. Blood vessel growth at later times relies on the process of angiogenesis, whereby new blood vessels are created from pre-existing ones. Two types of angiogenesis are known to occur: non-sprouting (intussusceptive and coalescent) and sprouting angiogenesis (see Figure~\ref{Fig1}). Intussusceptive, or splitting, angiogenesis arises when a capillary splits into two adjacent vessels. It occurs through the inward growth of ECs from opposite walls, which connect and partition the vessel into two luminal compartments, resulting in the formation of two new vessels \cite{Mentzer_2014} (see the left side of Figure~\ref{Fig1}). The discovery of intussusceptive angiogenesis dates back to the late 20th century \cite{Burri_2004}. Nonetheless, studying this phenomenon has been challenging due to difficulties visualising it through conventional light microscopy \cite{Mentzer_2014}. Further studies are needed to increase our understanding of the role and importance of intussusceptive angiogenesis in the formation of vascular networks \cite{Burri_2004,Mentzer_2014}. Recent studies of long-term light microscopy have shed light on an additional form of non-sprouting angiogenesis known as coalescent angiogenesis \cite{Nitzsche2022coalescent}. Unlike intussusceptive angiogenesis, this process involves the merging or coalescing of capillaries to enhance blood flow (see Figure~\ref{Fig1}, in the centre). Coalescent angiogenesis eliminates tissue islands that separate poorly perfused vessels, allowing them to fuse without disrupting blood circulation. This mechanism facilitates the transformation of an inefficient vascular network into a more efficient, hierarchical one \cite{Nitzsche2022coalescent}.

\begin{figure}
\begin{center}
\includegraphics[width=1\columnwidth]{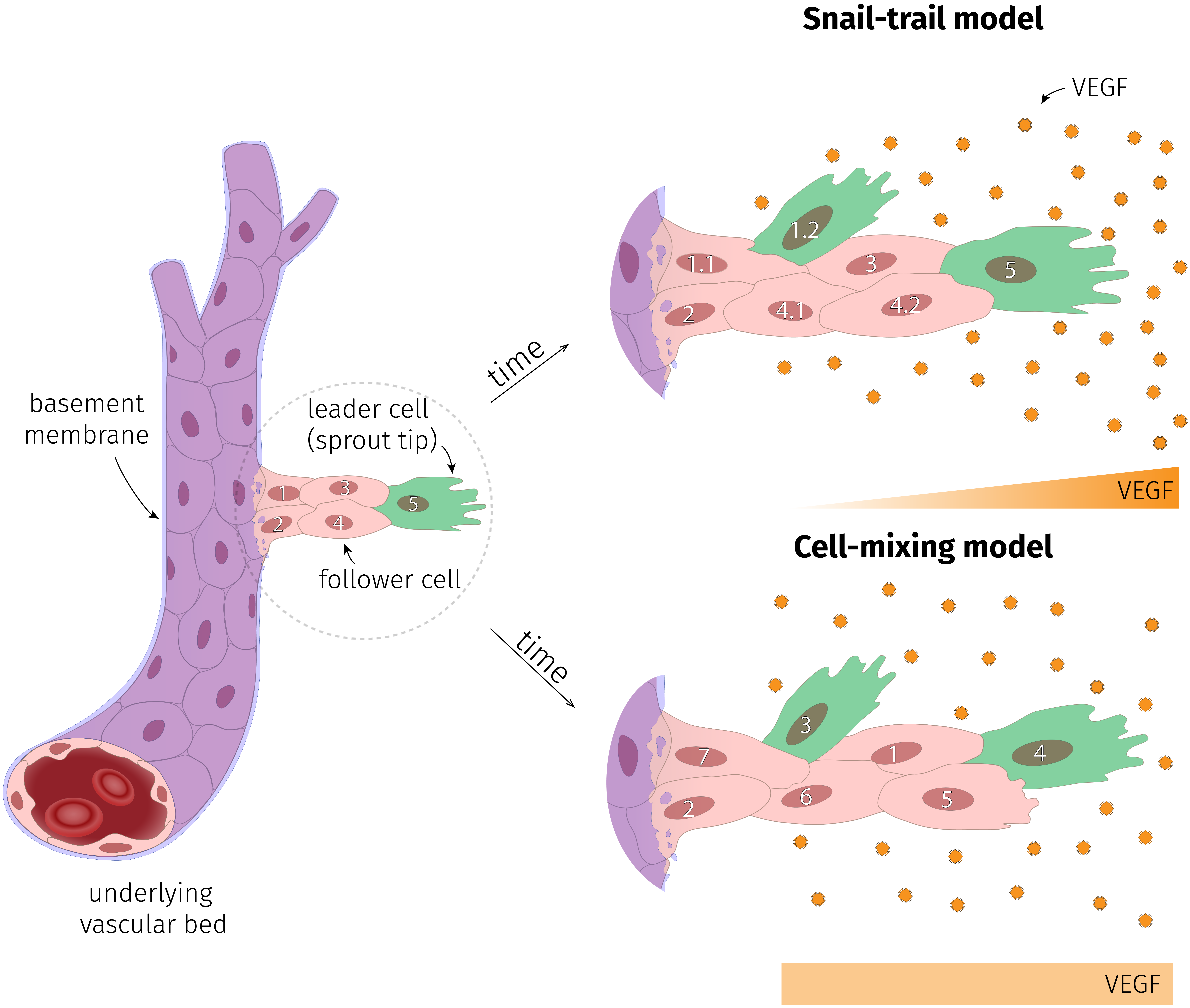}
\caption{\textbf{Snail-trail and cell-mixing models of sprouting angiogenesis.} The \emph{snail-trail~model} for vascular growth assumes that ECs situated at the forefront of sprouts assume a leadership role, guiding the elongation of sprouts in response to external cues such as spatial gradients of growth factors like VEGF. The remaining cells follow behind the leading cells and proliferate to replenish the elongating sprouts (in the cartoon for the snail-trail model, cell numbers with dots indicate daughter cells after proliferation). The leader-follower roles are determined according to the fixed cell positions within the growing sprout (i.e. cell order is invariably preserved). On the other hand, in the \emph{cell-mixing~model} of angiogenesis, ECs migrate from the underlying vascular bed where most of the cell proliferation occurs. Cell behaviours, or phenotypes, change dynamically via subcellular signalling in response to the microenvironment, which leads to cell mixing within elongating sprouts. Furthermore, although EC activation by VEGF is crucial in both models, the cell-mixing model does not require a spatial gradient in VEGF for vessel growth; initial vessel growth is migration-driven and can occur in spatially uniform concentrations of VEGF. However, in both models, a gradient in growth factors is needed to direct network expansion towards poorly oxygenated regions.}
\label{Fig2}
\end{center}
\end{figure}

Although the non-sprouting form of angiogenesis (intussusception and coalescence) can play a significant role in vascular growth and remodelling (especially during early embryonic development), sprouting angiogenesis is generally considered the dominant mode of vessel formation in adults \cite{Burri_2004,Nitzsche2022coalescent}. Thus, significant research efforts have been dedicated to investigating this type of blood vessel growth, which is characterised by sprouting of new vessels from pre-existing ones (see the right side of  Figure~\ref{Fig1}). Vascular endothelial growth factor (VEGF) is widely recognised as one of the most prominent signals that promote sprouting angiogenesis. The VEGF family comprises multiple isoforms whose importance in regulating angiogenesis varies depending on the context and specific biological process \cite{shibuya2011vascular}. VEGF-A, generally considered the most important and best-characterised isoform for angiogenesis, has several splice variants, including VEGF121, VEGF165, and VEGF189, which differ in their ability to bind to extracellular matrix components and their affinity for VEGF receptors \cite{shibuya2011vascular}. VEGF is typically secreted by poorly oxygenated (hypoxic) tissues. As it diffuses through the tissue, VEGF activates ECs that form the inner layer of blood vessels by binding to receptors on its cell membranes. Once activated, ECs break down the surrounding basement membrane and extend towards the hypoxic tissues, guided by local chemical and mechanical cues \cite{Phng_2009,Michaelis_2014,Senger_2011}. ECs can rapidly transition from quiescent to active migratory and proliferative states \cite{De_Bock_2013}. This rapid response facilitates the quick expansion of a new network of blood vessels needed to supply oxygen and nutrients to oxygen-deprived tissues. Angiogenic sprouting takes place in both health and disease. For example, insufficient angiogenesis has been observed in several pathological conditions, including retinopathy and cardiovascular diseases, whereas excessive vascular growth is characteristic of pathologies such as cancer and diabetic retinopathy \cite{Carmeliet_2005,Potente_2017}. Due to its importance in disease, understanding the mechanisms of angiogenesis has been an active area of research.

Under normal physiological conditions, the initial stages of vascular growth are characterized by expansion of a preliminary network consisting of numerous immature and poorly interconnected vessels \cite{Wietecha_2012}. Subsequently, through interactions with other cell types, such as pericytes, and in response to external stimuli such as blood flow, this network undergoes remodelling and stabilisation, ultimately giving rise to a functional vasculature \cite{Potente_2017,Geudens_2011,Mukwaya_2021,Franco_2015}. Disease can disrupt these processes, leading to the formation of abnormal vascular networks \cite{Carmeliet_2005,Potente_2017}. This review focuses on the early stages of physiological angiogenesis, during which initial branching networks form.

Until recently, mathematical modelling of angiogenesis has been dominated by a `follow-the-leader' paradigm, whereby ECs at the leading edge of growing sprouts sense external cues and direct sprout elongation towards their source.  At the same time, ECs that trail behind the leading cells ensure sprout integrity via proliferation. This model is known as the `\emph{snail-trail}' model of angiogenesis (see Figure~\ref{Fig2}). However, experimental findings have revealed that EC behaviour is more complex than suggested by the snail-trail assumption. ECs have been observed to overtake one another and exchange positions within growing sprouts \cite{Arima_2011,Jakobsson_2010}, a phenomenon known as \emph{cell mixing} or \emph{rearrangement} (see Figure~\ref{Fig2}).

\begin{figure}[h!]
\begin{center}
\includegraphics[width=0.75\columnwidth]{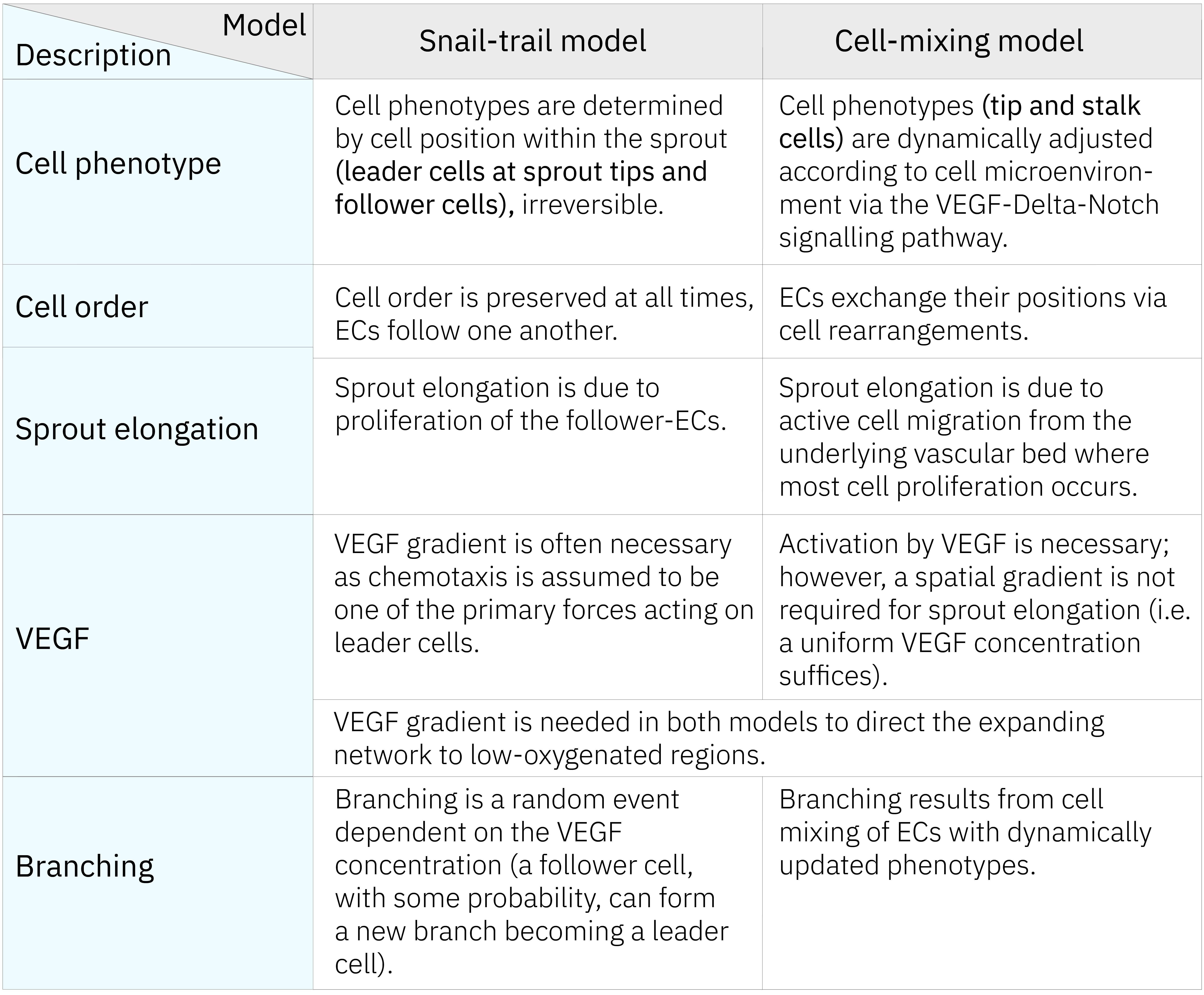}
\captionof{table}{\textbf{Comparison of characteristic aspects of the snail-trail (e.g. \cite{Pillay_2017}) and cell-mixing (e.g. \cite{Stepanova_2021}) models of angiogenic sprouting.}}
\label{Table1}
\end{center}
\end{figure}

One of the first experimental works demonstrating that ECs mix during angiogenic sprouting was by \cite{Arima_2011}. The authors designed a method to visualise the trajectories of individual ECs as they sprout from aortic ring assays embedded in a flat collagen matrix (\textit{in vitro}). By using time-lapse microscopy, they observed that ECs exhibit forward and backward motion within developing sprouts and that they can overtake one another. Analysis of their experimental data revealed that cell rearrangements contribute to sprout elongation, while only a small fraction of cells undergo proliferation on timescales from hours to days. These results indicate that during the early stages of angiogenesis (occurring within hours), sprout extension is primarily driven by cell migration (see Figure~\ref{Fig2}). These findings are further supported by experiments on retinal vascularisation in mice, which reveal that cell overtaking also takes place \textit{in vivo} \cite{Arima_2011}. Taken together, these observations provide strong evidence that the behaviour of ECs during angiogenesis is more complex than the snail-trail model suggests. Arima and coworkers also demonstrate that a small branching network can sprout from the underlying vasculature even when VEGF concentrations are uniform \cite{Arima_2011}. This further challenges a key assumption of the snail-trail model, which proposes that chemotactic migration up the spatial gradients of growth factors (e.g. VEGF) is one of the primary mechanisms driving sprout growth, even during early vessel formation. We note, however, that a gradient in growth factors is needed in both theoretical frameworks (snail-trail and cell-mixing models) to direct the expanding vascular network towards hypoxic tissues and ensure their successful vascularisation. We refer to Figure~\ref{Fig2} and Table~\ref{Table1} for more detailed descriptions of the snail-trail and cell-mixing models of sprouting angiogenesis.

Arima and colleagues demonstrated that EC rearrangement relies on subcellular signalling through the VEGF-Delta-Notch pathway \cite{Blanco_2012}. This finding aligns with conclusions obtained by \cite{Jakobsson_2010}, who employed mosaic sprouting assays to examine the ability of ECs with distinct gene expression profiles to compete for the leading position. Their \emph{in vitro} and \emph{in vivo} experiments revealed that ECs expressing higher levels of Delta are more likely to overtake their neighbours and assume the leadership role at the sprout tip than ECs with lower Delta levels.

While angiogenic sprouting has been extensively investigated \cite{Angulo_Urarte_2018,Arima_2011,Bentley_2014,De_Bock_2013,Cruys_2016,Franco_2015,Jakobsson_2010,Sugihara_2015,Ubezio_2016,Vega_2020}, the complex dynamics of individual ECs during angiogenesis are not fully understood. In particular, the mechanisms that drive EC rearrangements and their effect on the structure of the resulting vascular networks remain unclear \cite{Arima_2011,Jakobsson_2010,Angulo_Urarte_2018}.

Mathematical and computational modelling provides a convenient alternative to technically demanding and time-consuming experimental studies. However, most existing theoretical models of angiogenesis neglect phenotype-dependent behaviours of individual ECs, such as cell rearrangements, and instead use a coarser, snail-trail description of sprouting (see later in section~\ref{Section3}). Although mathematical and computational models of angiogenesis have been extensively reviewed (e.g. \cite{Scianna_2013,Vilanova_2017a,Mantzaris_2004,Heck_2015,Peirce_2008,Apeldoorn_2022}), this is the first review focusing on cell rearrangements during early angiogenic sprouting. In particular, we review snail-trail and cell-mixing models of angiogenesis (see also Figure~\ref{Fig2} and Table~\ref{Table1}) and discuss when the effects of cell rearrangements on vascular morphology cannot be neglected. This review draws upon, to a limited extent, research presented in the PhD dissertation \cite{stepanova2022mathematical}.

The structure of the review is as follows. Section~\ref{Section2} summarises biological processes occurring during physiological (healthy) angiogenesis. As such, in order to highlight the multiscale nature of this process, we review EC subcellular signalling specifying their phenotypes (section~\ref{Section21}), cell rearrangements (section~\ref{Section22}) and EC interactions with the surrounding environment, such as the extracellular matrix (section~\ref{Section23}). We also describe EC proliferation (section~\ref{Section24}) and the processes that occur during the later phases of angiogenesis, after the expansion of the initial vascular network (section~\ref{Section25}). Section~\ref{Section3} reviews the most-common theoretical approaches used to model angiogenesis within the snail-trail framework which we categorise according to their type (continuum models, section~\ref{Section31}, and discrete and hybrid models, section~\ref{Section32}). In section~\ref{Section4}, we describe in more detail several computational models of early sprouting which explicitly account for cell mixing. We conclude in section~\ref{Section5} by summarising the strengths and weaknesses of both the snail-trail and cell-mixing models of angiogenesis and proposing potential directions for future research in this area.

\section{Biological background} \label{Section2}

\subsection{Phenotype specification of endothelial
cells} \label{Section21}

\begin{figure}[h!]
\begin{center}
\includegraphics[width=0.91\columnwidth]{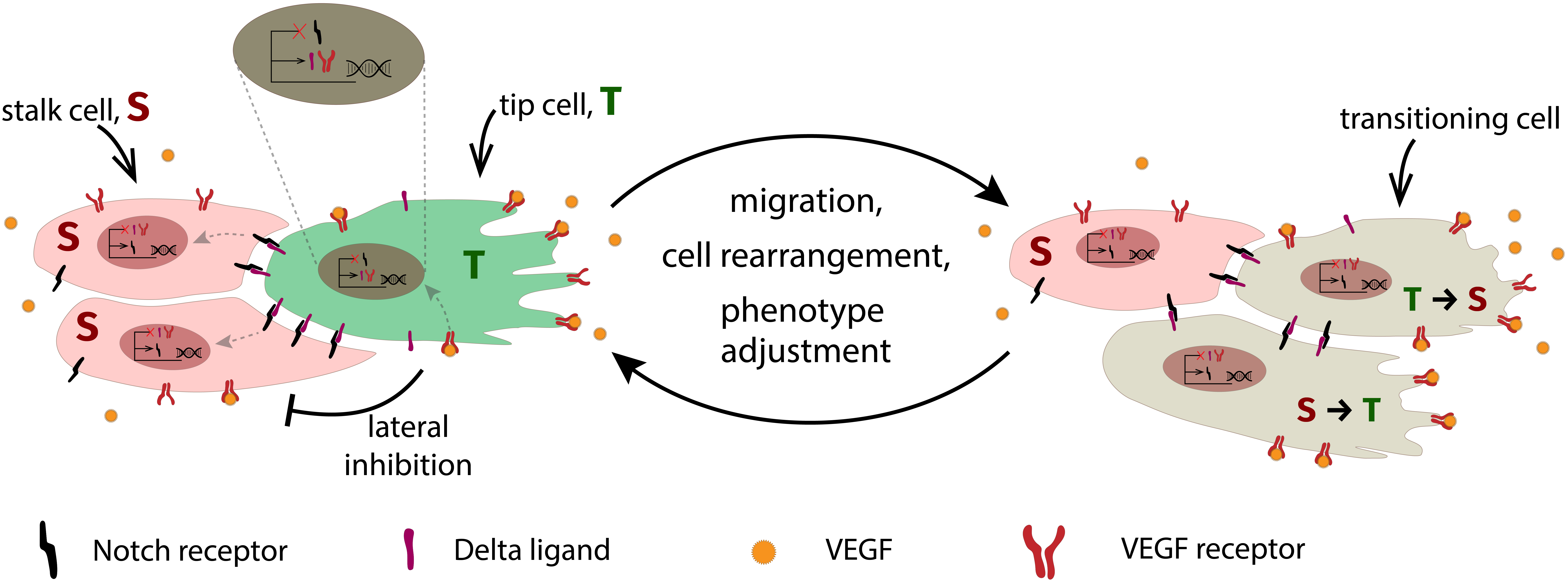}
\caption{\textbf{Phenotype specification of endothelial cells in
angiogenesis.} ECs rely heavily on the contact-dependent VEGF-Delta-Notch signalling pathway to determine their gene expression profiles. Tip cells (with high Delta/VEGF receptors and low Notch levels) are more polarised in shape and extend filopodia which enhance their ability to sense external cues and guide sprout elongation. By contrast, stalk cells (characterised by low Delta/VEGF receptors and high Notch levels) maintain vessel integrity and proliferate. Through Delta-Notch signalling, tip cells can inhibit their neighbours from acquiring the same phenotype (lateral inhibition mechanism) and ensure maintenance of an adequate tip-stalk cell ratio. It is hypothesised \cite{Jakobsson_2010,Arima_2011} that heterogeneous behaviours of exploratory tip and more passive stalk cells lead to cell mixing and, thus, to dynamic re-establishment of the gene expression patterns for each cell to maintain this tip-stalk cell ratio (i.e. ECs can transition between tip and stalk cell phenotypes). In this diagram, pink (green) represents stalk (tip) cells, while transitioning cells that adjust their gene expression are shown in pastel yellow on the right-hand side. Note that this schematic is a two-dimensional simplification of a three-dimensional cell arrangement within the sprout; instead of positioning themselves next to each other, ECs tend to crawl over one another, especially when overtaking (e.g. \cite{Sugihara_2015}).}
\label{Fig3}
\end{center}
\end{figure}

It is generally accepted that there are two main EC phenotypes during sprouting angiogenesis: migratory tip cells and proliferative stalk cells \cite{Blanco_2012}. However, this terminology needs to be clarified as tip and stalk cells are frequently used in snail-trail and cell-mixing frameworks but with different meanings (especially in the context of theoretical models, for example, \cite{Pillay_2017}). Within the snail-trail framework, phenotypes are associated with cell positions within the sprout. In this way, ECs at the leading positions of growing vessels are usually called tip cells (i.e. they are at the tip of the sprout); the rest of the cells trail behind the leading cells and are referred to as stalk cells. This is an idealised characterisation since cell position alone does not determine phenotype. Instead, cell phenotypes are characterised by specific gene expression profiles established via subcellular signalling pathways in accordance with extracellular cues perceived from a cell's local environment (Figure~\ref{Fig3}). The VEGF-Delta-Notch pathway is believed to be the dominant signalling pathway governing EC phenotype \cite{Blanco_2012,Gerhardt_2003,Hellstr_m_2007}. VEGF is one of the primary pro-angiogenic factors, while Delta and Notch act as transmembrane ligands and receptors, respectively. Their trans-binding capability enables intercellular communication, wherein a ligand on one cell binds to a receptor on another cell. Thus, this signalling pathway is contact-dependent. The VEGF-Delta-Notch pathway in the context of angiogenesis is known to produce a pattern of alternating cells with two distinct gene expression profiles: tip cells (characterised by high Delta and low Notch levels) and stalk cells (characterised by low Delta and high Notch levels). This subcellular signalling relies on lateral inhibition: tip cells repress their neighbours from acquiring a tip cell phenotype and drive them to become stalk cells (Figure~\ref{Fig3}). Since the cell at the leading position in a sprout has fewer neighbouring cells, it frequently acquires a tip cell phenotype leading to the misinterpretation that cell position determines phenotype. In order to avoid this confusion (see also Table~\ref{Table1}), we will refer to cell phenotypes (\textit{tip and stalk cells}) as specific gene expression profiles determined via subcellular signalling and to the \textit{leader (or sprout tip)} and \textit{follower cells} as the distinct roles (determined by cell position within the sprout) adopted within the snail-trail framework.

When activated by VEGF, ECs first acquire a tip cell
phenotype \cite{Phng_2009,Hellstr_m_2007}. Tip cells extend long thin membrane protrusions called filopodia that enable them to migrate actively. Since filopodia are enriched in VEGF receptors, they increase cell surface area and enable tip cells to guide sprout growth towards the source of VEGF. Tip cells also perform proteolysis; they secrete matrix metalloproteinases, a particular type of enzyme, that degrade the surrounding extracellular matrix (ECM) and create vascular guidance tunnels along which sprouts can elongate \cite{Phng_2009}. The second, stalk cell, phenotype is needed to maintain the growing sprouts' integrity and limit excessive exploratory behaviour by the tip cells.

High Delta levels in tip cells enable them to initiate the Notch signalling cascade in neighbouring ECs. Notch signalling inhibits expression of Delta and VEGF receptors in these ECs and, in turn, upregulates Notch, making neighbouring cells more susceptible to inhibition by Delta from the tip cell. Cells with high Notch, low Delta and few VEGF receptors become stalk cells \cite{Hellstr_m_2007,Blanco_2012}. Stalk cells migrate along the vascular guidance tunnels explored by the tip cells, and Notch signalling allows them to proliferate to maintain the integrity of the elongating sprouts. 

It is believed that distinct behaviours of the two cell phenotypes lead to cells exchanging their positions (i.e. cell microenvironment changes) and, then, to re-establishment of their gene expression profiles via VEGF-Delta-Notch signalling (Figure~\ref{Fig3}) \cite{Arima_2011,Jakobsson_2010}. This dynamic adjustment of cell phenotype maintains the tip-stalk cell ratio. The balance between the exploratory migration of tip cells and more quiescent stalk cell behaviour is vital for the formation of adequate vascular networks. This has been confirmed by \emph{in vitro} experiments in which Notch signalling was inhibited or completely blocked \cite{Jakobsson_2010}: in the absence of Notch signalling, all ECs adopted a tip cell phenotype, which led to excessive and aberrant sprouting and pathological network formation.

\subsection{Cell rearrangements} \label{Section22}

\begin{figure}
\begin{center}
\includegraphics[width=0.91\columnwidth]{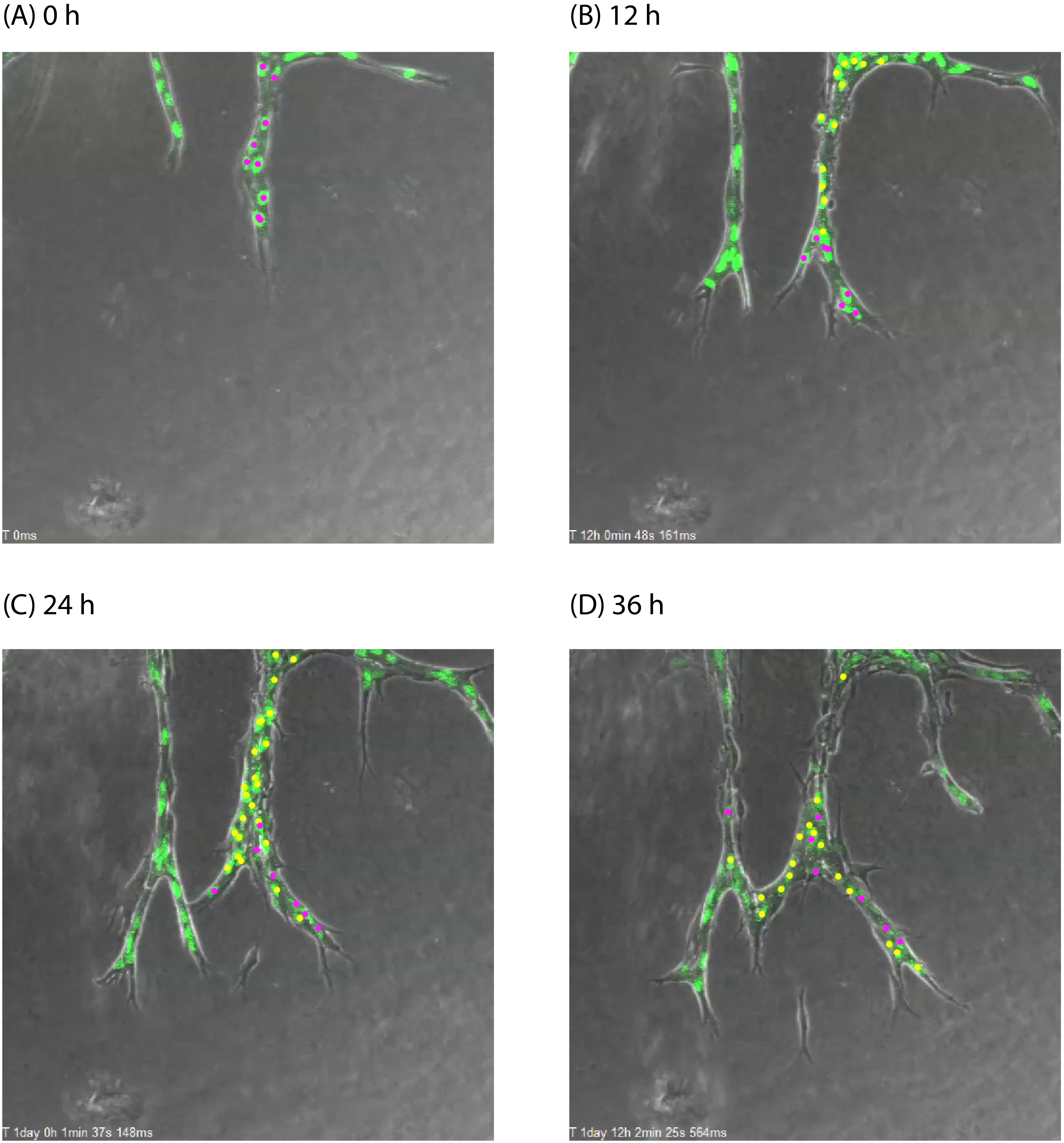}
\caption{\textbf{Cell mixing in angiogenesis.} The series of snapshots shows how angiogenic sprouts cultured \textit{in vitro} evolve over a time period of $36$ hours. Cell nuclei were selectively labelled with two distinct colours, magenta and yellow, to facilitate position tracking. Over time, as indicated in the title of each image panel, clusters of cells with the same colour gradually mix as cells overtake one another and shuffle within sprouts. These observations illustrate the phenomenon of \emph{cell rearrangement}, or \emph{cell mixing}, during the early stages of sprouting angiogenesis. \\
\small{Reprinted from Cell Reports, 13(9), K. Sugihara, K. Nishiyama, S. Fukuhara, A. Uemura, S. Arima, R. Kobayashi, A. Köhn-Luque, N. Mochizuki, T. Suda, H. Ogawa, H. Kurihara, ``Autonomy and non-autonomy of angiogenic cell movements revealed by experiment-driven mathematical modeling'', 1814-1827, Copyright (2015), with permission from Elsevier.}}
\label{Fig4}
\end{center}
\end{figure}

One of the first experimental models used to study tissue neovascularisation (and angiogenesis as its most common type) was the cornea (e.g. \cite{Folkman_1974,Gimbrone_1974,Muthukkaruppan_1979}). Normally avascular and transparent, corneas allow visualisation of angiogenic sprouting when stimulation by an implanted source of pro-angiogenic factors (frequently, a tumour implant) induces vessel outgrowth from the limbal region. The available experimental techniques at that time enabled the description of sprout outgrowth as a whole (visualisation of the behaviour of the individual ECs that comprise the vascular sprouts was not possible). Nonetheless, it was hypothesised that the sprout tips sensed the pro-angiogenic cues and directed sprout elongation towards their source \cite{Folkman_1974,Gimbrone_1974,Muthukkaruppan_1979}. The necessary supply of cells to the growing sprouts was attributed to proliferation of ECs that trailed behind the leading cells \cite{Folkman_1974,Gimbrone_1974,Muthukkaruppan_1979}. These experimental observations gave rise to the snail-trail model of angiogenesis. Numerous theoretical models of angiogenesis are based on the snail-trail hypothesis assumption (e.g. \cite{Balding_1985,Anderson_1998,Plank_2004,Pillay_2017}; see also reviews \cite{Chaplain_2000,Mantzaris_2004,Scianna_2013} and section~\ref{Section3}).

After several decades of active research, advances in visualisation techniques mean that it is now possible to zoom into growing vessels and characterise the behaviour of individual ECs. In contradiction to the snail-trail hypothesis, ECs were observed to actively exchange positions by crawling over one another \cite{Arima_2011,Jakobsson_2010}. Figure~\ref{Fig4} (reused from \cite{Sugihara_2015}; see also \cite{Arima_2011} for previous work of this research group) presents experimental evidence of cell mixing observed \textit{in vitro}. Sprouting from the aortic ring assay was induced by externally supplied VEGF. The small depth (compared to the other spatial dimensions) of the collagen substrate enabled the authors to track the trajectories of individual cells using confocal microscopy. To visualise cell rearrangements, EC nuclei were labelled with two different colours (magenta and yellow, as shown in Figure~\ref{Fig4}). Under the snail-trail hypothesis, the integrity of clusters of cells with the same colour label was expected to be preserved over time. However, the findings presented in Figure~\ref{Fig4} reveal that, in practice, cells actively exchange positions, leading to mixing of colour clusters.

\cite{Arima_2011,Jakobsson_2010} also showed that cell proliferation has a negligible effect on the sprout elongation on the timescale of hours. Cell division was minimal in the angiogenic front and was concentrated in the underlying vascular bed. They concluded that the growth of sprouts and expansion of the vascular network during the initial stages of angiogenic sprouting was dominated by cell migration and rearrangements, contrary to the proliferation-driven assumption of the snail-trail framework \cite{Arima_2011,Jakobsson_2010,Bentley_2014,Angulo_Urarte_2018}.

Cell mixing was shown to be related to subcellular
signalling \cite{Jakobsson_2010}. However, the precise mechanisms linking these processes remain unclear; it is not known whether the tip cell phenotype confers an advantage to an EC and allows it to occupy the leading position, or whether an arbitrary EC that finds itself at the leading position acquires the tip cell phenotype to perform its leader functions such as proteolysis and chemotactic guidance. Although it is acknowledged that the level of cell mixing influences the architecture of the growing vascular networks (e.g. reduced cell mixing leads to the formation of abnormal vascular networks with an excessive number of thin unfunctional vessels \cite{Angulo_Urarte_2018}), there is no clear mechanistic explanation of how cell rearrangements affect vasculature structure \cite{Angulo_Urarte_2018,Jakobsson_2010,Bentley_2009,Chen_2019}.

\subsection{Cell-ECM interactions} \label{Section23}

Descriptions of vascular growth must account for the microenvironment in which it occurs. We already mentioned how pro-angiogenic factors (such as VEGF) activate and direct sprout growth to poorly oxygenated tissues. Another key component of the microenvironment is the extracellular matrix (ECM). ECM is a mesh-like, dense, fibrous network composed of various collagens, fibronectin and other proteins \cite{Sottile_2004}. There is a dynamic coupling between the EC behaviour and the ECM; indeed, the structure and composition of the matrix (e.g. fibre orientation, density, pore size, and stiffness) control sprout growth, while EC migration leads to ECM remodelling \cite{Sottile_2004,Shamloo_2010,Du_2016}.

ECs migrate by forming adhesions with matrix fibres and crawling along them \cite{Reinhart_King_2005}. Thus, the alignment of ECM fibres influences the direction and speed of cell migration. In particular, it was reported that there is an optimal ECM density and stiffness for the formation of sprouts; soft or low-density substrates cannot support sprout formation (ECs migrate into the substrate in a scatter-like manner without forming sprouts), whereas highly stiff or dense matrices significantly limit sprout elongation in three-dimensional environments \cite{Shamloo_2010}. At the same time, EC migration remodels the ECM. We have already mentioned that tip cells use proteolysis to degrade the ECM and create space for sprout elongation \cite{Stratman_2009}. ECs also secrete new matrix components (such as collagen IV, fibronectin and various laminins \cite{Michaelis_2014}), which are needed to assemble the future basement membrane. It is a specialised type of ECM that surrounds blood vessels and limits EC migration by stabilising the newly generated capillaries \cite{Geudens_2011}. Another source of ECM remodelling is the mechanical force generated during EC migration. When ECs use their filopodia to pull on ECM fibres, they reallocate (i.e. bring closer) and reorient the fibres in the direction of sprout growth \cite{McCoy_2018,Sottile_2004}. As a result, bundles of aligned ECM fibres accumulate in the direction of the growing sprout and provide additional structural support to vessel elongation \cite{Kirkpatrick_2007}. 

\begin{figure}[h!]
\begin{center}
\includegraphics[width=0.91\columnwidth]{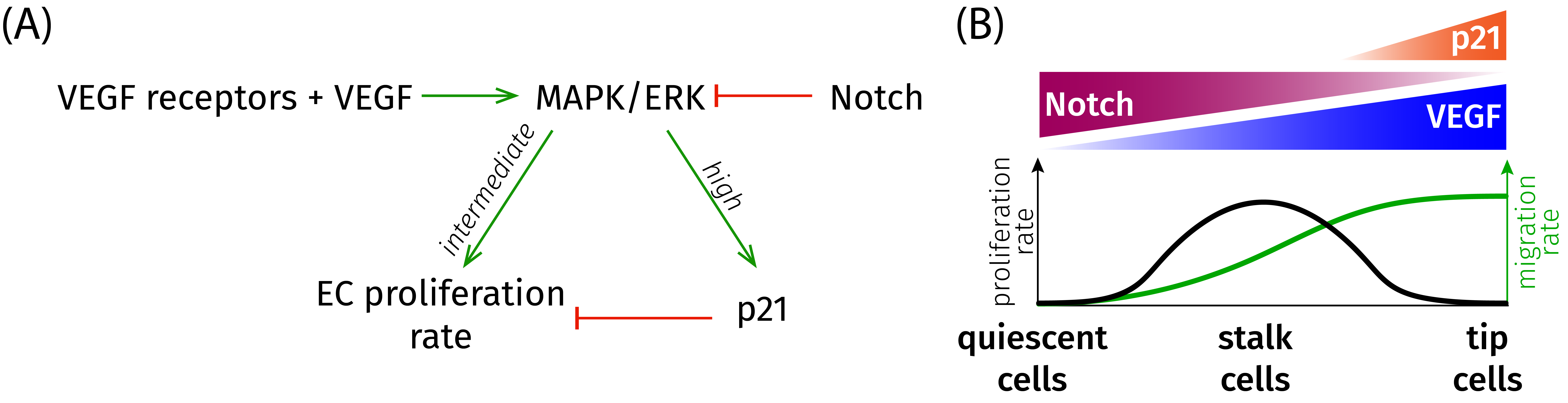}
\caption{\textbf{Endothelial cell cycle.} (A) A schematic of the signalling pathways involved in regulating EC proliferation. The MAPK/ERK signalling pathway (mitogen-activated protein kinases, MAPK, also referred to as extracellular signal-regulated kinases, ERK) plays a crucial role in regulating the G1/S transition of the cell cycle \cite{Lavoie_2020}. Activation of the MAPK/ERK pathway is induced by VEGF and inhibited by Notch signalling. The interaction between these two pathways drives stalk cell proliferation. Conversely, elevated MAPK/ERK activity in tip cells triggers the expression of a cell cycle inhibitor, p21. Here, the sharp green (blunt red) arrows indicate activation (inhibition). (B) As a result, the response of ECs to VEGF activation is characterised by a biphasic pattern, with a higher proliferation rate in stalk cells \cite{Pontes_Quero_2019}.}
\label{Fig5}
\end{center}
\end{figure}

\subsection{Cell proliferation and apoptosis} \label{Section24}

Cell proliferation occurs over a longer timescale, typically spanning 1-4 days \cite{Herz_2018,Zheng_2013,Form_1986,Frye_2002,Snead_1995,Anagnostou_1990}, than that associated with cell migration which typically spans minutes. Nonetheless, cell proliferation is crucial for adequate vessel growth and thickening \cite{Phng_2009}. In mature vessels, quiescent ECs maintain homeostasis with a low proliferation rate \cite{Eelen_2018}. At the onset of angiogenesis, ECs, activated by pro-angiogenic factors, adopt migratory tip and proliferative stalk cell phenotypes. While tip cells are characterised by low proliferation rates, stalks cells, particularly in the underlying vascular bed, actively proliferate \cite{Arima_2011,Jakobsson_2010}. This behaviour has been attributed to VEGF activation and its downstream signalling cascades \cite{Pontes_Quero_2019}. Optimal levels of VEGF signalling in stalk cells promote active proliferation. In contrast, higher VEGF activation in tip cells triggers the expression of a cell cycle inhibitor \cite{Pontes_Quero_2019} (refer to Figure~\ref{Fig5}). Thus, EC differentiation into tip and stalk cells via the VEGF-Delta-Notch signalling also regulates the EC cycle. External factors, including mechanical stimuli, can also regulate cell division \cite{Lamalice_2007,Lesman_2016}.

Activation of ECs by VEGF also plays a crucial role in maintaining their viability during the energetically demanding process of network expansion. Specifically, when VEGF binds to its cell surface receptors, it inhibits apoptosis and promotes cell survival \cite{Chavakis_2002}. Under physiological conditions, as the newly formed vascular network restores the oxygen supply to previously hypoxic tissues, levels of external VEGF decrease. This marks the transition from an angiogenic phase to a remodelling phase, where vessels with insufficient or turbulent blood flow regress and the vascular network matures \cite{Wietecha_2012}. In the absence of the protective effects of VEGF signalling, ECs undergo apoptosis, eliminating excessive branches that compromise the functionality of the vasculature. In pathological conditions, the anti-angiogenic switch is dysregulated (e.g. due to excessive VEGF secretion by tumour cells) \cite{Carmeliet_2005}. This dysregulation leads to a disorganized vascular network with excessive branching and thin, tortuous vessels incapable of effectively sustaining blood flow \cite{Carmeliet_2005,Potente_2017}.

\subsection{Later stages of angiogenesis} \label{Section25}

After the initial expansion of an immature vascular network, the subsequent stages of angiogenesis involve vascular remodelling and stabilisation. Vascular remodelling encompasses various changes in the structure and organization of the blood vessels, including alterations in diameter (dilation and constriction), length, branching patterns, and cellular composition. A specific aspect of remodelling is vessel pruning, which involves the selective elimination or regression of redundant blood vessels. This process occurs when ECs detect cues indicating insufficient or turbulent blood flow. In response, ECs may undergo apoptosis \cite{Wietecha_2012, Watson_2017} or migrate towards high-flow vessels \cite{Franco_2015} to improve nutrient delivery. The remaining vessels experience increased blood flow, leading to the activation of Kruppel-like factor 2 (KLF2). KLF2 promotes EC quiescence by slowing cellular metabolism and inhibiting proliferation \cite{Zecchin_2017}. Furthermore, during vessel stabilisation, ECs recruit pericytes and smooth muscle cells, which attach to the outer surface of the blood vessels. These supportive cells play a crucial role in strengthening EC junctions, ensuring tight connections between cells, and facilitating the assembly of a basal lamina around the vessels \cite{Bergers_2005}.

\section{Theoretical models of angiogenesis without cell mixing} \label{Section3}

\begin{figure}
\begin{center}
\includegraphics[width=0.98\columnwidth]{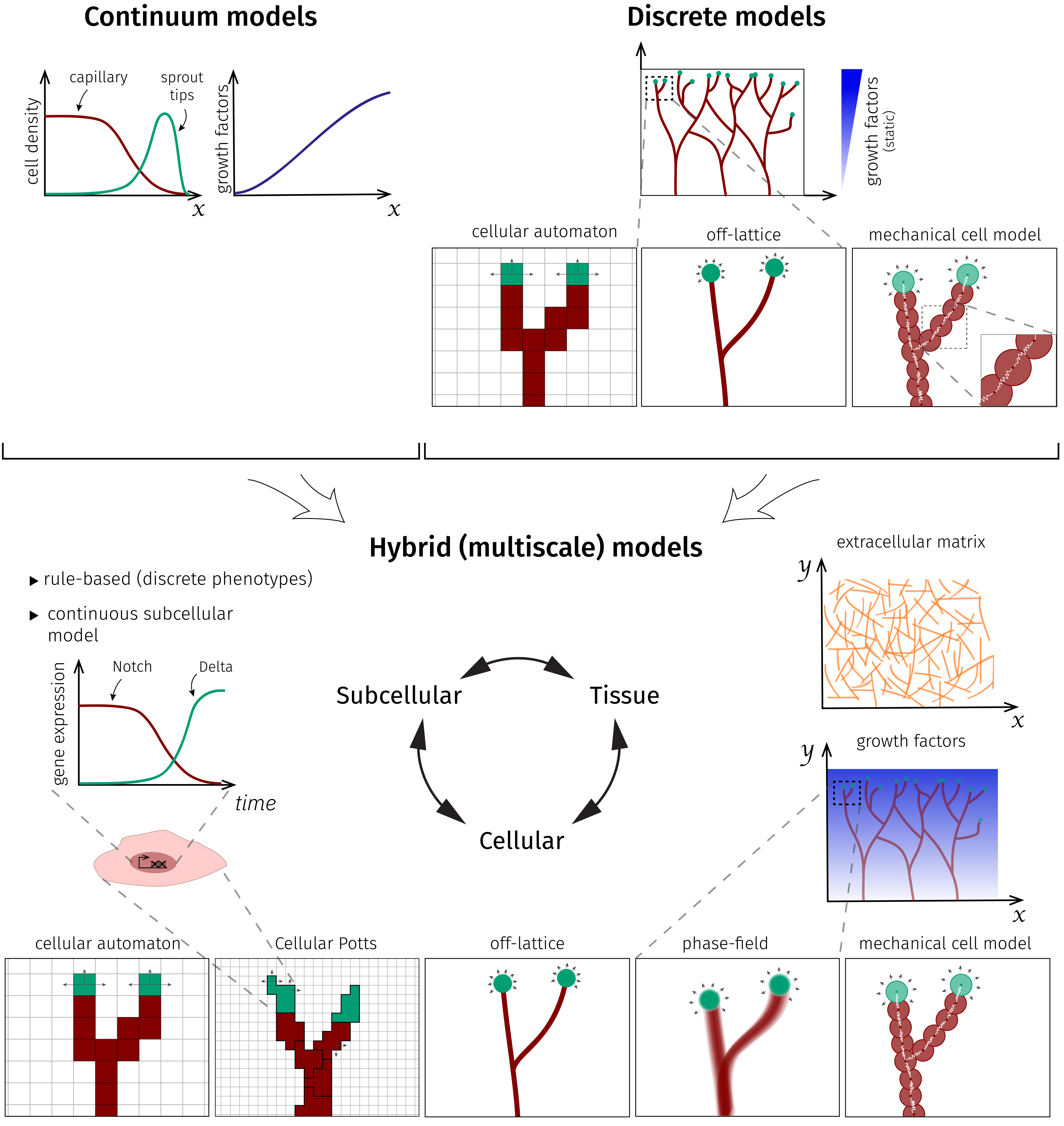}
\caption{\textbf{A cartoon illustrating different techniques used to model angiogenesis within the snail-trail framework.} Early theoretical models described angiogenic sprouting deterministically in terms of ordinary/partial differential equations. Later, in order to reproduce the branching structure of vascular networks, discrete models treated sprout tips as individual agents which can migrate in on-, off-lattice fashion or, for example, due to mechanical cell-cell interactions (e.g. node-based models). The capillary is then represented by the track left behind by the sprout tips. Hybrid models combine deterministic and discrete techniques. The cellular scale is often modelled discretely. In addition to the already mentioned cellular automaton, off-lattice and mechanical models, frequently used modelling techniques include Cellular Potts Models and phase-field models in which sprout tips are treated as discrete agents. On the other hand, tissue scale variables (e.g. growth factors) are typically modelled deterministically. This makes hybrid models naturally multiscale. Some models can also include discrete or continuous descriptions of the extracellular matrix at the tissue scale and cell phenotypes at the subcellular scale. Here, leading sprout tips are shown in green, and the trailing capillary (follower cells) are shown in red.}
\label{Fig6}
\end{center}
\end{figure}

Numerous theoretical models have been proposed to investigate different aspects of sprouting angiogenesis in physiological and pathological contexts. In limited space, it is not possible to review all of these models. Here, we highlight representative works which illustrate different modelling approaches that are most widely used in the field (see also Figure~\ref{Fig6} for illustration). Existing models, which generally assume a snail-trail mechanism and do not account for cell mixing, are reviewed in section~\ref{Section31} for continuum models and section \ref{Section32} for discrete and hybrid models. We categorise discrete and hybrid models according to the approach used to model EC dynamics (e.g. cellular automaton, off-lattice models, etc.). We also mention if model simulations were compared to experimental data (if nothing is stated, then no comparison to experiments was carried out). For alternative reviews of the existing literature on mathematical and computational (snail-trail) models of angiogenesis, we refer the reader to the reviews \cite{Scianna_2013,Vilanova_2017a,Mantzaris_2004,Heck_2015,Peirce_2008,Apeldoorn_2022}. Other reviews, with a more specialised focus, include the work by Flegg et al. on wound-healing and tumour-induced angiogenesis \cite{Flegg_2020} (includes a timeline indicating significant modelling works in this context), reviews from a systems biology perspective \cite{Logsdon_2013,Zhang_2021}, and multiscale models of angiogenesis \cite{Qutub_2009}. The angiogenesis reviews by \cite{Peirce_2012,Bentley_2013} also highlight the importance of interdisciplinary collaborations between experimentalists and theoreticians. In section~\ref{Section4}, we present a more detailed description of several computational models that account for cell rearrangements and dynamic phenotype specification. 

\subsection{Continuum models} \label{Section31}

Angiogenesis in the context of tumour growth motivated the development of early theoretical models of vascular growth. These models were typically formulated in terms of partial differential equations for the density of expanding blood vessels and pro-angiogenic stimuli such as VEGF that stimulate sprout outgrowth (e.g. \cite{Deakin_1976,Balding_1985,Chaplain_1993,Byrne_1995,Chaplain_1996}). Many of the early models focused on reproducing the \emph{brush-border effect} (i.e. the experimental observation of increased branching and blood vessel density closer to the source of the growth factors \cite{Muthukkaruppan_1982}) and identifying conditions for successful neovascularisation of hypoxic tissue (sensitivity to angiogenic stimuli, EC proliferation). Extensions of these models of angiogenic sprouting (e.g. \cite{Orme_1996,Orme_1997}) included the ability of ECs to move up spatial gradients of adhesive substances (fibronectin, different types of collagens) secreted by other ECs, a phenomenon termed \emph{haptotaxis}. \cite{Levine_2001} formulated a model to study the onset of sprout growth when ECs, activated by pro-angiogenic stimuli, degrade the surrounding basal lamina via proteolysis and initiate sprouting towards the source of the angiogenic cues. 

Later continuum models investigated vascular growth in a broader range of settings. For example, \cite{Aubert_2011} studied angiogenesis in the context of murine retina vascularisation. Their model incorporates the migration of astrocytes from the optic nerve within the retina, which plays a crucial role in guiding the expansion of the vascular network through the secretion of growth factors. Model simulations in \cite{Aubert_2011} were shown to be in good agreement with experimental data on the temporal advancement of the front of migrating ECs and astrocytes. In separate work, \cite{Flegg_2009} developed a model to study angiogenesis during wound healing. They investigated how hyperbaric oxygen therapy, which involves delivering increased oxygen levels to wounds, can promote the healing of chronic lesions. \cite{Connor_2015} extracted quantitative data from images of corneal vasculature to parametrise early snail-trail continuum models reported in \cite{Balding_1985,Byrne_1996}.

The models discussed above neglect the influence that the structure of the underlying ECM has on EC behaviour. It has been observed that EC migration contributes to ECM deformation and remodelling \cite{Senger_2011}. As a result of this matrix reorganisation, neighbouring ECs may experience passive movement. \cite{Manoussaki_1996} developed a mechanical continuum model of vasculogenesis to investigate how these cell-ECM interactions can give rise to the characteristic vascular patterns observed \textit{in vitro}. \cite{Holmes_2000} used this framework to study angiogenesis; their model incorporates the reorganisation of the ECM resulting from traction forces due to EC migration, as described by \cite{Manoussaki_1996}, along with sprout growth dynamics similar to those proposed in \cite{Chaplain_1993,Orme_1997}. The primary objective of their study was to examine how angiogenesis is affected by the mechanical properties of the ECM and its interactions with ECs.

In the context of vasculogenesis, vascular architecture was also resolved by assuming that ECs secrete a chemoattractant which leads to the recruitment of adjacent ECs. This assumption was sufficient to generate a connected vascular network from initially dispersed ECs in deterministic models of \textit{de novo} vascular growth \cite{gamba2003percolation,serini2003modeling} (reviewed in \cite{Scianna_2013}).

\subsection{Discrete and hybrid models} \label{Section32}

The deterministic approaches mentioned above, with the exception of the vasculogenesis models proposed by \cite{Manoussaki_1996,gamba2003percolation,serini2003modeling}, do not capture the complex vascular architecture of developing networks, which is crucial for assessing blood flow dynamics, nutrient delivery, and treatment efficacy in cases of pathological angiogenesis. To address this limitation, new models have emerged that view ECs as individual entities, enabling the tracking of their trajectories and the blood vessels they generate. These models can be classified as either \emph{discrete}, assuming a static external environment, or \emph{hybrid}, combining a continuum approach to describe the dynamics of the cell microenvironment with a cell-based EC model. Due to the significant influence of the extracellular environment on cellular behaviour, only a few existing models are `purely' discrete, such as those by \cite{Sleeman_2002,Pillay_2017}. The hybrid modelling approach has been more widely employed in angiogenesis research and can simulate vascular growth in a manner that closely resembles experimentally observed branching networks. Moreover, hybrid models can facilitate comparison to experiments since metrics quantifying vascular architecture (e.g. branching angles, vessel length, or metrics provided by more advanced techniques such as Topological Data Analysis, \cite{Nardini2021topological}) and statistics on individual cell behaviour (such as cell trajectories and velocities) can be extracted from simulations. 

The hybrid modelling approach offers the flexibility to formulate detailed models which are capable of resolving vascular architecture and complex cell behaviours, such as cell rearrangements. Here, we focus on hybrid modelling techniques used to model angiogenesis within the snail-trail framework. Models which incorporate cell rearrangements are reviewed in section~\ref{Section4}.

\subsubsection{Cellular automata} \label{Section321}

Cellular automata are mathematical models in which cells (agents) can move on a fixed grid (see Figure~\ref{Fig6}), and their positions and states (e.g. phenotype) are updated over discrete time steps based on a set of predetermined rules. These rules determine how the state of a cell evolves and may depend on the states of its neighbours and its microenvironment (e.g. growth factors).

One of the pioneering on-lattice models for sprout growth was developed by \cite{Anderson_1998}. They initially formulated a 2D continuum angiogenesis model and subsequently discretised it to create a rule-based cellular automaton that simulates the growth of individual sprouts. This approach involves treating EC as discrete entities within a lattice, where the probabilities of migration in specific directions depend on the growth factor levels. The model also incorporates rule-based processes for branching and vessel fusion (anastomosis). This model has been extended in numerous ways; in \cite{Chaplain_2000}, the model was extended to 3D; McDougall and coworkers subsequently included a model for blood flow and drug delivery in 2D \cite{McDougall_2002} and 3D \cite{St_phanou_2005}. Later works by this research group also accounted for the dynamic coupling of sprouting with simultaneous vascular remodelling due to blood flow \cite{McDougall_2006} and the effects of pericyte coverage on vessel stabilisation \cite{McDougall_2010}. The study conducted by \cite{Bartha_2006} also explored the process of vascular remodelling in the context of tumour growth. More detailed reviews of these works can be found in \cite{Chaplain_2006} and other models of vascular remodelling in \cite{Rieger_2015}. Another extension of the work of \cite{Anderson_1998} includes a model of corneal angiogenesis developed by \cite{Watson_2012}; they coupled angiogenic sprouting with an equation for astrocytes (secrete growth factors that promote vascular growth). The authors also compared the simulated vascular networks to experimental images of retinal vascularisation. Their simulations are in good agreement with the experimental data, capturing the temporal expansion of the retinal vascular plexus and the position of the astrocyte front.

Vascular dynamics (vascular remodelling and angiogenesis) have been considered in several multiscale models of tumour growth. Alarc\'{o}n and colleagues used a multiscale hybrid approach to investigate vascular adaptation in tumour environments \cite{Alarc_n_2003,Alarc_n_2005}. Their model accounts for interactions between stromal and cancerous cells together with the VEGF dynamics and the evolution of the underlying vasculature in terms of blood flow, remodelling, and oxygenation. This model was later extended to include vessel dematuration and increased vessel permeability caused by the loss of the basement membrane under high VEGF stimulation induced by tumours and used to evaluate the efficacy of different cancer treatments \cite{Alarc_n_2006,Byrne_2006}. The model by Alarc\'{o}n et al. was also extended by \cite{Owen_2009} to account for changes in vascular architecture due to angiogenesis and pruning of vessels caused by low shear forces (insufficient blood flow). \cite{Perfahl_2011} used this modelling framework to study vascular tumour growth and its response to treatment in three-dimensional environments. \cite{Macklin_2009} presented another work highlighting the importance of coupling vascular adaptation to tumour growth. Their multiscale model combined a new detailed description of tumour invasion with vascular growth and remodelling as in \cite{Anderson_1998,McDougall_2006}.

\cite{Sleeman_2002} introduced an alternative approach in which they proposed a stochastic model for EC migration at sprout tips, employing a reinforced random walk description. The migration probabilities in their model were influenced by chemotactic and haptotactic stimuli, enabling the investigation of the interplay between these external cues in angiogenesis regulation. Another stochastic model for angiogenic sprouting was developed by \cite{Spill_2014} using a mesoscopic approach. They derived transition rates for the stochastic processes of EC migration and proliferation, allowing for the derivation of a corresponding mean-field description of vascular growth. This mean-field model facilitated the identification of conditions under which the continuum description provides a reliable approximation of the original stochastic model. \cite{Pillay_2017} also employed a discrete-to-continuum approach to obtain macroscopic models of angiogenesis from individual cell dynamics. They formulated a 2D cellular automaton for sprouting within the snail-trail framework, from which they derived a one-dimensional continuum model to describe macroscopic sprout growth. \cite{Martinson_2020} extended this work by deriving a two-dimensional continuum model that accurately captures the individual EC behaviour observed in the discrete model proposed by \cite{Pillay_2017}.

Another example of a 3D cellular automaton includes the work of \cite{Norton_2016}. One of their objectives was to gain insights into how changes in the relative contributions of EC proliferation and migration influence the branching pattern of the developing vascular network.

\subsubsection{Cellular Potts Models without cell mixing} \label{Section322}

The Cellular Potts Model (CPM) is an extension of the cellular automaton that introduces cell shape representation \cite{Graner_1992,Glazier_1993} (see Figure~\ref{Fig6} for illustration). This is accomplished by associating a cell with a cluster of small lattice sites and employing a Monte Carlo algorithm for iterative displacements of the cell boundary. This results in the dynamic evolution of cell shape and position, driven by the intrinsic properties of the cell and its interaction with the surrounding microenvironment. The CPM modelling approach has been effectively applied to investigate vasculogenesis and angiogenesis (see below for the references of representative works). CPM models of vascular growth typically adopt a hybrid methodology, where deterministic treatment is applied to tissue-scale variables (such as external stimuli and ECM) while incorporating stochastic dynamics for individual cells. \cite{Bauer_2007,Bauer_2009} developed one of the first CPM models of tumour-induced angiogenesis, focusing on the impact of the ECM as a barrier for sprout extension (although the model is limited to two dimensions).

The CPM framework has also been employed to investigate the mechanisms underlying the formation of vascular honeycomb patterns observed in \textit{in vitro} vascular growth. One mechanism, known as the contact-inhibition model, demonstrates that the reduced chemotactic sensitivity in the parts of EC membranes that are in contact with neighbouring ECs can give rise to the generation of characteristic vascular networks observed in vasculogenesis and angiogenesis \cite{Merks_2008}. An alternative explanation was proposed by \cite{van_Oers_2014}. They presented a CPM that incorporates the ECM deformation due to contractile forces exerted by cells. The mechanical strains within the ECM, in turn, modulate EC behaviours through a phenomenon known as durotaxis, where cell sensitivity to substrate rigidity guides their movement. This purely mechanical framework successfully reproduced the formation of interconnected vascular networks in the context of vasculogenesis and angiogenesis. The role of ECM and its remodelling in regulating vascular growth was also investigated by \cite{Daub_2013}. Their simulation results demonstrated that chemotaxis alone is capable of supporting the growth of linear sprouts without branching. However, the formation of branching networks requires the combined influence of chemotaxis, proteolysis, cell proliferation, and haptokinesis (the ability of cells to adjust their migration speed in response to ECM concentrations).

\cite{Shirinifard_2009} developed a three-dimensional CPM of vascular growth in tumour environments. In particular, they used their multicellular model to investigate how the evolution of the surrounding vasculature influences tumour morphology. 

\subsubsection{Off-lattice models} \label{Section323}

Off-lattice models, as the name suggests, do not rely on a predefined grid or lattice structure; cells, represented as individual particles, are allowed to move freely within the model domain (see Figure~\ref{Fig6}), and their positions and states evolve according to rules based on physical principles, or the specific behaviours being modelled. One of the early off-lattice models that address the growth of individual capillaries was proposed by \cite{Stokes_1991}. In this model, capillary sprout tips are considered distinct entities, and their velocities are determined by a set of stochastic differential equations. The density of each vessel, represented by the trajectory of sprout tips, evolves over time, as governed by a separate ordinary differential equation. Another model, presented by \cite{Plank_2004}, offers an alternative approach, where the migration of each sprout tip is characterised by its speed and direction. Reorientation, i.e. changes in the direction of motion, is modelled as a random walk on a unit circle.

An off-lattice model of cornea vascularisation was proposed by \cite{Tong_2001}. In this model, the growth of blood vessels was induced by a pro-angiogenic stimulus (basic fibroblast growth factor, bFGF) released from a pellet implanted into the cornea. The model accounts for sprouting from limbal vessels at the cornea circumference in response to the bFGF stimulation. Building upon this work, \cite{Harrington_2007} extended the model to investigate vascular growth in the cornea in the presence of angiogenesis inhibitors. 

In more recent work, diverse techniques have been used in order to develop more detailed but computationally feasible models of angiogenesis. For example, \cite{Sun_2005} used the finite element method to reduce the computational cost of simulations of their hybrid multiscale model of angiogenesis. \cite{Milde_2008} formulated the first three-dimensional hybrid model of sprouting, which explicitly incorporated the effects of ECM composition on the distributions of matrix-bound and soluble VEGF.

\subsubsection{Phase-field models} \label{Section324}

In the context of angiogenesis, the phase-field method combines an off-lattice approach for modelling individual sprout tips with a continuous representation of the capillary density given by the phase field. The phase-field variable indicates the presence or absence of blood vessels at different locations within the tissue. Its time evolution is governed by a system of partial differential equations, capturing the growth of blood vessels in response to the surrounding microenvironment and sprout tip migration (based on the snail-trail assumption). The dynamics of discrete sprout tips are coupled with the phase-field description of the trailing capillaries by setting the phase-field variable to specific fixed values at the positions of sprout tips.

One of the first angiogenesis models formulated using the phase-field methodology was proposed by \cite{Travasso_2011}. Their model incorporates deterministic migration of sprout tips up spatial gradients of VEGF secreted by hypoxic cells within the tissue domain. \cite{Vilanova_2014} extended this model by introducing noise into the migration of sprout tips. They viewed the direction of EC movement as a stochastic variable, following a random walk on a unit circle, as proposed by \cite{Plank_2004}. Simulations conducted in 2D and 3D dimensions demonstrated that the resulting vascular networks have lower connectivity in three dimensions. This emphasises the importance of anastomosis (vessel fusion) for the formation of functional vascular networks since poorly connected networks are unable to deliver an adequate supply of oxygen and nutrients. The process of anastomosis was further studied by \cite{Vilanova_2017}, who accounted for the ability of sprout tip filopodia to detect neighbouring vessels and VEGF concentrations. This mechanism enhanced the formation of connections between vessels and facilitated vascular regrowth along the empty collagen sleeves of regressed vessels. This approach allowed the authors to simulate the scenario of network regression due to inhibition of activation by growth factors and the subsequent regrowth of vasculature once inhibition was removed. Capillary densities at each stage of this numerical experiment (prior to inhibition, during the regression phase and the following regrowth) were closely aligned with data from \textit{in vivo} experiments. Alternative mechanisms of anastomosis have been explored by \cite{Moreira_Soares_2018}. Their simulation results suggest that the secretion of pro-angiogenic factors by hypoxic cells has a more significant impact on the branching patterns of growing networks than either EC division or the migration speed of sprout tips. The phase-field models developed by \cite{Travasso_2011,Vilanova_2017} have also been used to study vascular tumour growth in \cite{Xu_2016}.

\subsubsection{Mechanical cell models} \label{Section325}

Mechanical models represent sprouts as a collection (chain) of interconnected cell-agents. Each cell is identified with specific mechanical properties, such as stiffness or contractility, and is connected to its neighbours by links or springs that represent mechanical interactions between cells (see Figure~\ref{Fig6}). Mechanical models can account for forces associated with chemotaxis, persistence in elongation direction and cell-cell interaction that are frequently included in other modelling approaches for angiogenesis. This approach also allows for the inclusion of additional angiogenesis-specific forces, such as those generated by tension within developing sprouts. Vessel tension plays an important role in shaping vascular network morphology by influencing branching angles, as was shown in \cite{Secomb2013angiogenesis}.

One model of angiogenesis which accounts for mechanical cell-cell interactions is a model for the vascularisation of cornea proposed by \cite{Jackson_2010}. They used a viscoelastic model for the elongation and migration of sprout tips. The trailing stalk cells follow behind and proliferate in response to VEGF stimulation. A subcellular model, formulated as a system of differential equations within each cell, describes cell maturation, which promotes quiescence and vessel stabilisation in regions proximal to the growing vascular networks (i.e. EC proliferation is localised to the angiogenic front, where active sprouting occurs). An alternative approach was proposed by \cite{Perfahl_2017}. In their off-lattice model, ECs are viewed as spheres connected by linearly elastic elements to neighbouring cells within the same sprout. ECs migrate and divide in 3D in response to external forces due to chemotaxis, stretch and persistence. In addition, they introduced glyphs, a visualisation tool to summarise several properties of vascular morphologies simultaneously, and demonstrated how the same properties can be generated from experimental images. This facilitates comparison between \textit{in silico} and biological vascular networks. \cite{Phillips_2020} presented an agent-based model of angiogenic sprouting in tumour environments. The model accounts for sprout growth stimulated by VEGF and regulated by the mechanical forces arising from EC interactions with other cells (neighbouring ECs and cancer cells) and the surrounding ECM. The model also accounts for EC phenotype transitions (VEGF-dependent). 

\vspace{10pt}

Many existing computational models of angiogenesis were developed using custom software. Unfortunately, open-source code is not readily available for the majority of these models. This poses a challenge in the field as cross-comparison of different models and reproducibility of the reported results are frequently difficult to achieve. Microvessel Chaste, developed by \cite{Grogan_2017}, addresses this issue by offering an open-access library to simulate vascular growth in a variety of contexts (e.g. tumour invasion, angiogenesis, blood flow).

As the brief review presented above shows, many models of angiogenesis are formulated within the snail-trail framework. We now review alternative computational models (continuum models do not allow to resolve such a level of detail) of sprouting angiogenesis that account for the cell mixing phenomenon.

\section{Computational models with cell mixing} \label{Section4}

\subsection{The memAgent Spring Model (MSM)} \label{Section41}

\begin{figure}
\centering
\includegraphics[width =\linewidth]{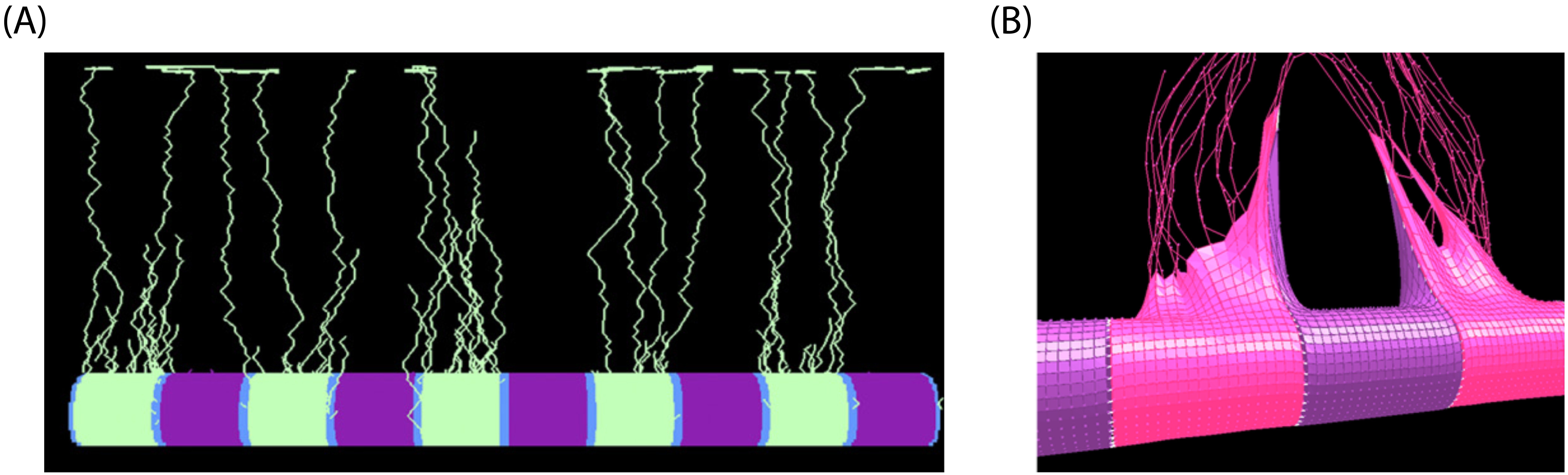}
\caption{\textbf{Representative simulations of the memAgent Spring Model by \cite{Bentley_2008,Bentley_2009}.} (A) Stabilisation of a pattern of ECs with alternating phenotypes (salt-and-pepper pattern) in simulations of the memAgent Model in an environment with linearly increasing, in vertical direction, VEGF \cite{Bentley_2008}. Stalk (tip) cells with low (high) levels of activated VEGF receptors are shown in purple (green); intercellular junctions are shown in blue. (B) Fusion of two tip cells (in pink) migrating up a spatial gradient of VEGF simulated using the memAgent Spring Model from \cite{Bentley_2009}. Stalk cells are shown in purple. \\ \small{(A) Reprinted from Journal of Theoretical Biology, 250(1), K. Bentley, H. Gerhardt, P. A. Bates, ``Agent-based simulation of notch-mediated tip cell selection in angiogenic sprout initialisation'',  25-36, Copyright (2008), with permission from Elsevier. (B) Reused from PLoS Computational Biology 5(10), K. Bentley, G. Mariggi, H. Gerhardt, P. A. Bates, ``Tipping the balance: Robustness of tip cell selection, migration and fusion in angiogenesis'',  e1000549, Copyright (2009), with permission from the publisher under a Creative Commons Attribution License.}}
\label{Fig7}
\end{figure}

In a series of articles \cite{Bentley_2008,Bentley_2009,Bentley_2014}, Bentley and colleagues developed a novel \textit{in silico} approach to investigate EC communication during angiogenesis. In their memAgent Model, cells are represented by a collection of memAgents, which correspond to small sections of the cell membrane. The memAgents are connected by linear elastic springs to mimic cell shape. This representation enabled them to investigate EC crosstalk within the fixed geometry of a blood vessel segment. The communication between cells was assumed to occur via the VEGF-Delta-Notch signalling pathway which was further facilitated by positive feedback resulting from filopodia extension (ECs activated by VEGF extend filopodia with VEGF receptors which increases the surface area for VEGF capture by the cell).

In their first memAgent Model, \cite{Bentley_2008} considered fixed cell shapes and investigated how filopodia extension impacts VEGF-induced cell phenotype specification to form the pattern of alternating stalk and tip cells (known as the `salt-and-pepper' pattern; see Figure~\ref{Fig7}(A) for representative simulations). Their findings suggest that filopodia can accelerate the establishment of the salt-and-pepper pattern. However, under high (pathological) VEGF levels, ECs exhibited synchronous behaviour in which their phenotype oscillated in time (all ECs adopt the same fate, stalk or tip cell). This behaviour can be attributed to saturation of the VEGF receptors; here, the extension of filopodia does not confer a competitive advantage for ECs to inhibit neighbouring cells and, as a consequence, all cells adopt the same fate. Later work by \cite{Bentley_2009} extends this framework to incorporate elongation of the EC membrane up the VEGF gradient and filopodia-guided anastomosis (memAgent Spring Model, MSM). They used the extended model to investigate the effects of sprout fusion (anastomosis) on the stability of EC phenotypes in diverse VEGF environments (see Figure~\ref{Fig7}(B)).

Further extensions of the MSM \cite{Bentley_2014} incorporate cell shuffling within the fixed tubular geometry to investigate the impact of cell rearrangements on early angiogenesis. Simulations of their \textit{in silico} model predicted that differential cell-cell adhesion (dependent on EC gene expression profiles) can give rise to cell shuffling. They demonstrated that two mechanisms are needed to generate cell rearrangements; one mechanism was based on the observation that VEGF activation reduces cell adhesion (via VE-cadherin endocytosis), the second on the observation that Notch signalling inhibits cell polarisation and, thus, the formation of junctional protrusions in the direction of cell migration. Both mechanisms favour tip cell migration (characterised by high VEGF and low Notch signalling). These computational modelling predictions were then validated experimentally \cite{Bentley_2014}. Given the significant computational demands associated with the MSM, simulations were performed using a simplified setup involving 10 ECs shuffling within a cylindrical domain. The authors comment that it is unclear how cell shuffling affects vascular morphology and that further MSM extensions are needed to investigate this.

The MSM has also been extended to mimic specific experimental setups to study various aspects of EC behaviour, such as the competitive ability of ECs to occupy the leading cell position \cite{Jakobsson_2010}, the EC division \cite{Costa_2016}, the impact of cell metabolism on cell mixing in angiogenesis \cite{Cruys_2016}, and the ability of EC filopodia to sense their environment and stimulate ECs to alter their subcellular signalling accordingly, rather than passively receive cues from the environment \cite{Zakirov_2021}.

\subsection{Cellular Potts Models with cell mixing} \label{Section42}

We now revisit the CPM approach to review in more detail two such models which account for cell rearrangements. 

\begin{figure}[h!]
\centering
\includegraphics[width = 0.9\linewidth]{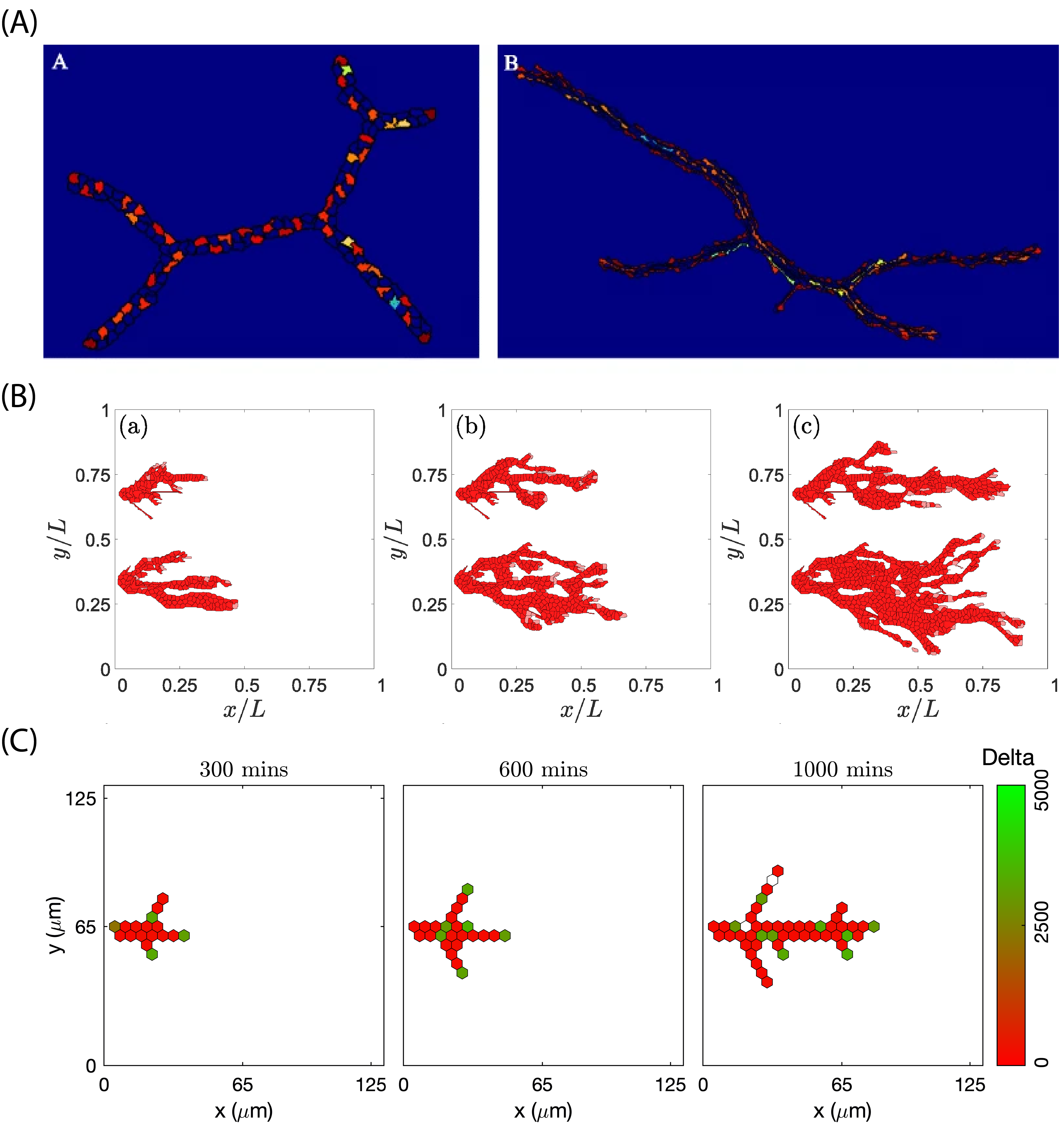}
\caption{\textbf{Representative simulations of Cellular Potts Models with cell mixing by \cite{Boas_2015,Vega_2020} and the multiscale model by \cite{Stepanova_2021}.} (A) Two CPMs of vascular growth (contact inhibition and cell elongation models; for more details, see \cite{Boas_2015}) used to investigate cell rearrangements in angiogenesis. Stalk (tip) cells with low (high) levels of Delta are shown in blue (red). (B) A CPM of angiogenic sprouting proposed by \cite{Vega_2020} to investigate the interplay between two Notch ligands, Delta and Jagged. Stalk (tip) cells with low (high) levels of activated VEGF receptors are shown in red (pink). (C) A multiscale hybrid model of early angiogenesis developed by \cite{Stepanova_2021} to investigate the effects of cell rearrangements on vascular architecture. Stalk (tip) cells with low (high) levels of Delta are shown in red (green). \\ \small{(A) Reprinted from BMC Systems Biology, 9, S. E. Boas, R. M. Merks, ``Tip cell overtaking occurs as a side effect of sprouting in computational models of angiogenesis'',  1-17, Copyright (2015), with permission from the publisher under a Creative Commons Attribution License. (B) Reused from PLoS Computational Biology 16(1), R. Vega R, M. Carretero, R. D. Travasso, L. L. Bonilla, ``Notch signaling and taxis mechanisms regulate early stage angiogenesis: A mathematical and computational model'',  e1006919, Copyright (2020), with permission from the publisher under a Creative Commons Attribution License. (C) Reused from PLoS Computational Biology 17(1), D. Stepanova, H. M. Byrne, P. K. Maini, Alarc\'{o}n, ``A multiscale model of complex endothelial cell dynamics in early angiogenesis'',  e1008055, Copyright (2021), with permission from the publisher under a Creative Commons Attribution License.}}
\label{Fig8}
\end{figure}

Motivated by the cell mixing results of \cite{Arima_2011,Jakobsson_2010}, Boas \& Merks employed the CPM approach to study this phenomenon within small vascular networks \cite{Boas_2015} (see Figure~\ref{Fig8}(A)). The advantage of the CPM framework is that these models approximate cell shapes, and cell trajectories can be tracked (since CPM lattice sites are small compared to the characteristic cell size). The CPM framework accounts for intrinsic noise, which naturally leads to ante- and retrograde movement of ECs within sprouts and cell overtaking. In this way, Boas \& Merks demonstrated that cell mixing occurred spontaneously in their simulations without accounting for differential behaviours of tip and stalk cells.

The authors also investigated whether subcellular signalling can fine-tune cell mixing. They incorporated a subcellular model based on ordinary differential equations of the VEGF-Delta-Notch signalling and considered two scenarios for the behaviour of the stalk and tip cells: (i) reduced adhesion for tip cells and (ii) lower sensitivity of tip cells to the external cues they perceive from other ECs. In order to incorporate these behaviours into the CPM framework, they assumed that the subcellular model produced discrete phenotypes: ECs acquired a tip/stalk phenotype when their Notch activity passed a predetermined threshold value (high/low Notch corresponds to stalk/tip cells). This model was then used to reproduce experimental results from \cite{Jakobsson_2010}.

In \cite{Jakobsson_2010}, the ability of two cell lines (wild type, WT, and mutant EC less sensitive to VEGF) to compete for the leading cell position was quantified (mosaic sprouting assay). The simulation results of Boas \& Merks agreed with experimental data from \cite{Jakobsson_2010}, with WT cells occupying leading cell positions more often than mutant ECs. Jakobsson et al. proposed that cell mixing regulated by VEGF-Delta-Notch signalling ensures the strongest (`optimally equipped') cell can shuffle up to the sprout tip \cite{Jakobsson_2010}. Since in the model of Boas \& Merks, cell mixing occurs irrespective of cell signalling, they proposed an alternative explanation for the mosaic assay experiments \cite{Boas_2015}. They noticed that due to their reduced sensitivity to VEGF, mutant cells less frequently acquired tip cell phenotype. Thus, the chances of finding a mutant tip cell at the leading position were lower (i.e. the observed distribution of WT and mutant cells at leading positions in sprouts was not due to cell overtaking). This interpretation suggested that cell mixing is a byproduct of sprouting and has no functional role. Therefore, Boas \& Merks proposed that the VEGF-Delta-Notch signalling plays a protective role, ensuring that there is always a tip cell in the leading position of a sprout when cell overtaking occurs spontaneously \cite{Boas_2015}.

The CPM approach by Boas \& Merks offers a flexible framework for investigating cell rearrangements. However, we note that, to our knowledge, the assumption that ECs produce a chemoattractant cue which attracts other ECs has not been confirmed experimentally. Another limitation of their model is that cell migration is restricted to two dimensions: in practice, cells may crawl over each other when overtaking (e.g. \cite{Sugihara_2015}), and this can influence the dynamics of cell mixing.

\cite{Vega_2020} developed another CPM of angiogenic sprouting that allows for cell mixing. However, investigating cell rearrangements was not the main focus of this work; the authors aimed to uncover the role of Jagged, another ligand for Notch, in the subcellular signalling of ECs (see Figure~\ref{Fig8}(B)). Previous studies had only examined the competition between the two Notch ligands, Delta and Jagged, in fixed geometries. Therefore, Vega et al. sought to understand how these subcellular dynamics influence sprouting when cells migrate and proliferate. Although the authors comment that cell overtaking occurs in their model, they did not explore it further. In addition to subcellular signalling, this model accounts for the mechanical strains generated by the cells in the ECM, which can guide cellular migration and anastomosis.

\subsection{A hybrid multiscale approach} \label{Section43}

Motivated by the cell mixing experiments from \cite{Arima_2011,Jakobsson_2010}, we investigated cell rearrangements in the context of early angiogenesis, when sprout elongation is migration-driven (and cell proliferation is negligible \cite{Arima_2011,Jakobsson_2010,Stepanova_2021}; see Figure~\ref{Fig8}(C) for illustration). We sought also to investigate the role of the ECM as a scaffold which guides sprout formation and cell migration (see section~\ref{Section23}). We formulated our model in a multiscale framework, accounting for VEGF-Delta-Notch signalling at the subcellular scale, cell migration
at the cellular scale and ECM remodelling due to sprouting at the tissue scale. Cell signalling was formulated as a stochastic gene regulatory network to account for the continuum of cell gene expression and its adjustments to the cell microenvironment (i.e. phenotype transitions). We implemented cell migration as a persistent random walk on a hexagonal lattice (tracking only the positions of cell nuclei), which is biased by the structure and composition of the ECM. Finally, matrix remodelling occurs upon cell migration (ECs re-align matrix fibres, secrete enzymes that degrade the ECM and form a basement membrane around the sprout). Cell overtaking is implemented as in a leapfrog model: the positions of cell nuclei on the lattice are exchanged when cells overtake each other. Coupling between the scales is incorporated by assuming that cell behaviour depends on its phenotype. Thus, gene expression patterns regulate cell migration, overtaking, and matrix remodelling.

After calibration and validation of the model with experimental data from \cite{Arima_2011,Sugihara_2015,Jakobsson_2010,Shamloo_2010}, we generated vascular networks for the different cell lines used in \cite{Jakobsson_2010} (WT and mutant cells with lower/higher sensitivity to VEGF). Using a number of metrics to quantify the branching patterns of networks composed of ECs from different lineages, we showed how changes in subcellular signalling can affect the morphology of simulated vascular networks. Since none of the cellular behaviours in our model were rule-based, we asked how individual cell signalling is translated into the macroscopic network morphology and whether it can be through cell rearrangements. We quantified cell mixing as the normalised mean distance that two neighbouring ECs travel away from each other during a fixed time period. We found that mixing levels for networks formed by mutant cells (both less and more sensitive to VEGF than WT) were lower than for WT. This suggested that cell subcellular signalling influences the resulting levels of cell mixing. At the same time, the reduced cell rearrangement in mutant ECs correlated with changes in the balance between sprout extension and branching in the generated vascular networks. These modelling results led us to propose that cell rearrangement (regulated by VEGF-Delta-Notch signalling) contributes to maintaining a balance between linear growth and lateral expansion of growing vessels in the following way \cite{Stepanova_2021}:

\begin{figure}[t]
\centering
\includegraphics[width=0.95\columnwidth]{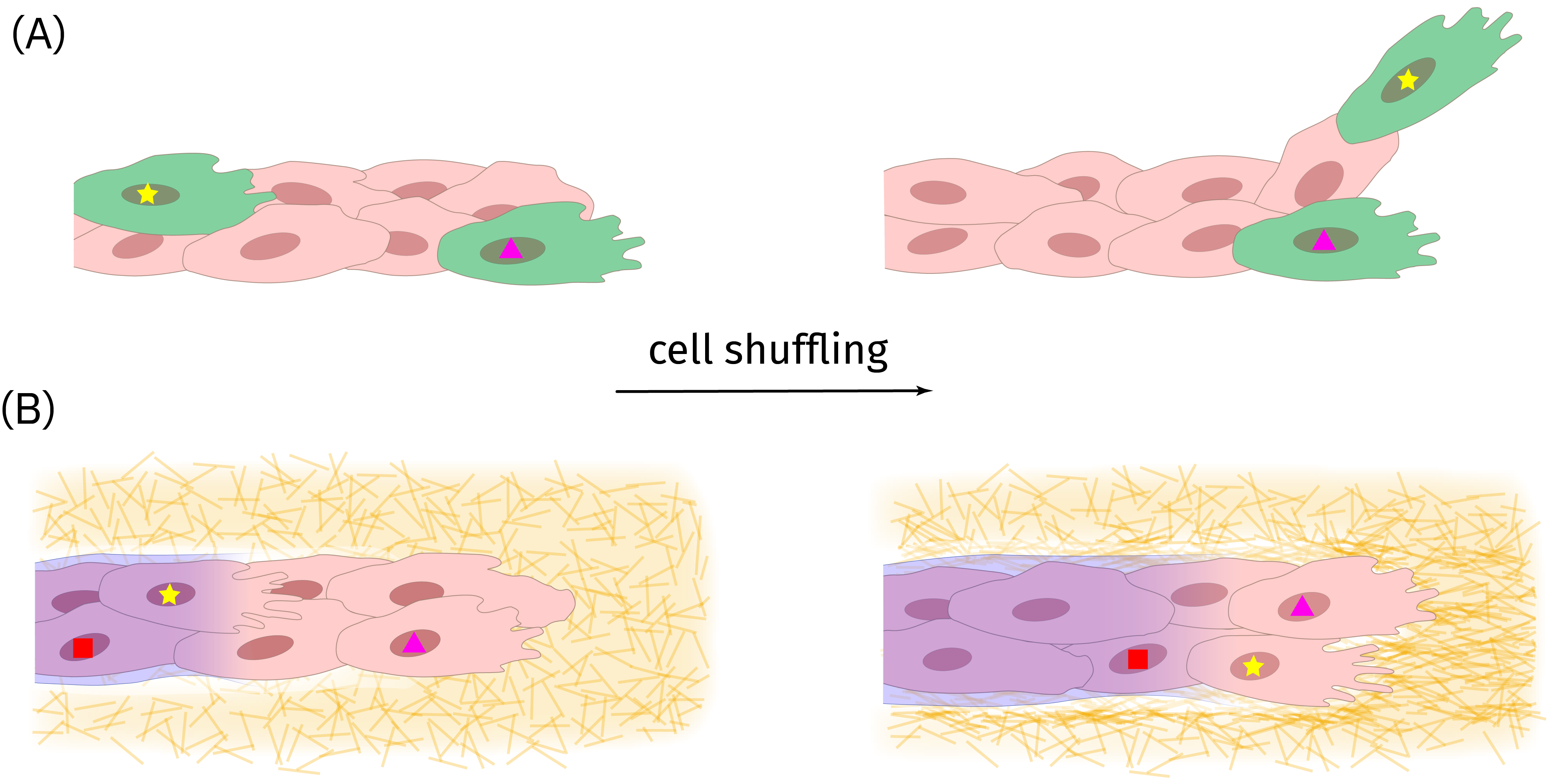}
\caption{\textbf{Schematic diagrams illustrating the role of cell mixing during early angiogenesis proposed by \cite{Stepanova_2021}.} (A) EC migration occurs on a faster timescale than changes in gene expression. As a result, a tip cell (labelled by a yellow star) can migrate to a different position within a vessel and initiate a new sprout before being inhibited by a tip cell in its vicinity (a magenta triangle indicates its nucleus). (B) Cell shuffling contributes to the remodelling of the surrounding ECM, such as the re-alignment of its fibrils (in orange), ECM degradation to facilitate the formation of guidance tunnels around sprouts, and the formation of future basal lamina (in blue). These changes in ECM structure and composition contribute to vessel stabilisation. Here, labels of different shapes were used to distinguish cells (phenotypes not shown).} \label{Fig9}
\end{figure}

\begin{itemize} \setlength\itemsep{0pt} 
\item Since changes in EC phenotype occur over a longer timescale than cell migration, exploratory tip cells can migrate, by shuffling, to new positions within a vessel and may create a new sprout before they are inhibited by neighbouring ECs (contribution to branching during early expansion; see Figure~\ref{Fig9}(A)). 
\item Cell rearrangement implies constant cell migration, which contributes to ECM remodelling that, in turn, guides vascular growth and promotes vessel stability (contribution to vessel stability at longer times; see Figure~\ref{Fig9}(B)).
\end{itemize}

While the above hypothesis on the role of cell rearrangements is yet to be tested experimentally, the mixing measure can be computed from experimental data which tracks cell trajectories. Comparing values of the mixing metric extracted from experimental and synthetic data represents the first step towards understanding whether our model predictions offer a plausible explanation for cell rearrangements in angiogenesis. We also note that the hexagonal lattice used in our model constrains the branching angles in the generated vascular networks: this feature must be improved in further studies by extending our model to incorporate lattice-free cell migration.

\section{Discussion} \label{Section5}

Angiogenesis plays a crucial role in a range of physiological and pathological processes. It is essential for embryonic development, wound healing, and tissue regeneration. Angiogenesis also contributes to tumour growth and metastasis, as well as in the progression of diseases such as cardiovascular disorders and retinopathies \cite{Potente_2017}. While half a century of research on angiogenic sprouting has greatly increased our knowledge of the mechanisms behind blood vessel formation, much remains to be understood. Theoretical modelling serves as a valuable tool in furthering our understanding of the complex biological interactions and mechanisms involved in vessel growth. Modelling complex systems necessarily involves making simplifications in a manner that is similar to the way in which experiments may be conducted in a controlled and idealised \textit{in vitro} environment in order to better comprehend a specific aspect of angiogenesis \textit{in vivo}. At the same time, it is crucial that models accurately reflect the essential characteristics of the process they aim to represent. Most existing theoretical models of angiogenesis (see section~\ref{Section3}) are formulated within the snail-trail framework of fixed leader-follower roles of ECs during sprouting. Recent experiments that track individual cells during sprouting reveal that cells do not simply follow each other but instead rearrange within growing blood vessels (cell mixing) \cite{Jakobsson_2010,Arima_2011,Sugihara_2015}. This evidence challenges the validity of the snail-trail assumption and raises questions about whether it is an appropriate simplification to use in theoretical models.

Here, we have reviewed our current understanding of the process of EC rearrangements during sprouting angiogenesis. We also examined various modelling approaches that can be used to describe sprout growth neglecting (snail-trail framework) or accounting for this complex EC behaviour (cell-mixing framework). Although we focussed on the differences between snail-trail and cell-mixing models of sprouting, we believe that, from a modelling standpoint, these frameworks are not mutually exclusive; rather, they offer different perspectives of sprouting angiogenesis, at different scales. At timescales from hours to days \cite{Arima_2011,Sugihara_2015}, the dynamics of individual cells are dominant and thus, cell mixing should be taken into account. However, on longer timescales, when considering the dynamics of a sprout as a whole, experimental results suggest that a leading tip cell can always be found to guide the sprout's growth, even though the identity of this cell may change over time. In this context, the snail-trail model is sufficient to describe the sprout dynamics.

The computational models discussed in section~\ref{Section4} \cite{Bentley_2014,Boas_2015,Vega_2020,Stepanova_2021} represent the initial efforts to account for cell mixing when modelling angiogenic sprouting. There are many ways in which these models could be extended to investigate the functional role of cell rearrangements on vascular morphology and angiogenesis in general. We suggest three key processes that future models should include. One fundamental component is the subcellular signalling that defines EC phenotypes and their behaviours during sprouting. Secondly, such models should account for off-lattice cell migration (i.e. EC movement should not be constrained by a lattice or fixed geometry), which allows quantifying of the architecture of simulated vascular networks. Lastly, the role of the ECM in providing physical scaffolding for sprout extension and ECM remodelling by migrating ECs suggest that cell-matrix interactions (e.g. proteolysis, fibre re-alignment) should also be incorporated. Since each of these three processes acts on a different scale - subcellular, cellular, and tissue scales, respectively - a hybrid multiscale modelling framework should be used to formulate cell-mixing models of angiogenesis. Furthermore, to validate these potential models, a more accurate quantification of cell mixing behaviour is needed than is currently available. We hope that more light will be shed on the effects of cell rearrangements during early angiogenesis by combining experimental and theoretical modelling approaches.

It is worth noting that collective behaviour, such as cell rearrangements, is not exclusive to angiogenesis and is observed across different biological scales, from single cells and bacterial colonies to animal groups \cite{Buttenschon_2020}. Theoretical models can be a powerful tool to investigate and identify the cues that drive collective migration and rearrangement. For instance, modelling studies have been used to identify key cues that govern the collective behaviour of neural crest cell migration \cite{Giniunaite_2020,Giniunaite_2020b,Shellard_2020}, cancer cell invasion \cite{George_2017,Stichel_2017}, bacterial swarming \cite{Peruani_2012,Jeckel_2019} and collective animal behaviours \cite{Giardina_2008}. We believe that theoretical modelling can be employed to bridge the gap between the microscopic and macroscopic scales and advance our understanding of how collective behaviours drive the emergence of macroscopic features such as vascular network morphology. Identifying the hallmarks of collective movement and rearrangement in various biological systems, including angiogenesis, could help uncover fundamental principles that govern the emergence of complex multicellular systems.

In conclusion, the field of angiogenic sprouting still faces many unresolved questions that motivate the development of theoretical models. Our understanding of cell mixing and its role in angiogenesis remains limited, including whether a cell that assumes tip cell phenotype through cell signalling has an advantage when it competes for the leading position in an angiogenic sprout \cite{Bentley_2014,Stepanova_2021} or whether occupying the leading position confers an advantage on a cell to adopt the tip cell phenotype and carry out its functions (as suggested in \cite{Boas_2015}). While it is generally accepted that ECs adopt tip and stalk cell phenotypes during angiogenesis, the mechanisms by which cells transition between these phenotypes is still uncertain. It is unclear whether a cell's behaviour changes continuously during this transition or whether transitions between characteristic behaviours occur in a switch-like manner (discrete phenotypes). Additionally, the existence of a hybrid cell phenotype (displays characteristics of both tip and stalk cells) \cite{Boareto_2015,Venkatraman_2016}, particularly under pathological conditions (explored in the sprouting model by \cite{Vega_2020}), remains an open question. ECs with a hybrid phenotype are thought to behave partially like exploratory tip cells while maintaining the ability to proliferate like stalk cells, resulting in an excessive number of thin vessels that are unable to sustain adequate blood flow. Another question which requires further investigation is non-local cell-cell interactions mediated by the ECM remodelling. While it is known that ECM remodelling by ECs influences vascular growth, it remains uncertain whether these cell-ECM feedbacks always promote vascular expansion and how they are altered in pathological conditions. Mathematical and computational modelling is a powerful tool which can help us to find answers to these questions and many others to advance our understanding of angiogenesis and its mechanisms.

\section*{Acknowledgements}

The authors apologise for the numerous studies left out due to the space constraints of this review.

\section*{Conflict of interest}

The authors have declared no conflicts of interest for this article.

\section*{Data availability statement}

Data sharing is not applicable to this article as no new data were created or analysed in this study.


\end{document}